\documentclass[aps,pra,superscriptaddress,longbibliography,twocolumn,10pt]{revtex4-2}
\usepackage{graphicx}
\usepackage{amsmath,amssymb,mathtools,bbm,bbold,braket,hyperref,xcolor,multirow,setspace,dsfont,microtype}
\usepackage[utf8]{inputenc}
\usepackage[english]{babel}
\usepackage{pgf}
\hypersetup{colorlinks=true, linkcolor=blue, citecolor=blue, urlcolor=blue}
\usepackage{comment}

\usepackage{mwe}
\usepackage{enumitem}
\graphicspath{{./FigsPRL}}

\DeclarePairedDelimiterX{\matrixelement}[3]{\langle}{\rangle}{#1 \lvert #2 \rvert #3}
\DeclarePairedDelimiter{\expval}{\langle}{\rangle}
\newcommand{\ii}{i}

\newcommand{\me}{e}

\newcommand{\prlsection}[1]{\noindent \emph{#1}.---}

\newcommand{\steady}{stationary}

\newcommand{\mycomment}[1]{}

\newcommand{\Gammafunc}{\Tilde{\Gamma}}
\newcommand{\commutator}[2]{\left[#1,#2\right]}

\makeatletter
\renewcommand{\fnum@figure}{FIG.~S\thefigure}
\makeatother

\begin{document}


\title{Symbolic Quantum-Trajectory Method for Multichannel Dicke Superradiance}

\author{Raphael Holzinger}
\email{raphael.holzinger@uibk.ac.at}
\affiliation{Institute for Theoretical Physics, University of Innsbruck, Technikerstr. 21a, A-6020 Innsbruck, Austria}
\affiliation{Department of Physics, Harvard University, Cambridge, Massachusetts 02138, USA}
\author{Nico S. Bassler}
\affiliation{TU Darmstadt, Institute for Applied Physics, Hochschulstra\ss e 4A, D-64289 Darmstadt, Germany}
\author{Julian Lyne}
\affiliation{Max Planck Institute for the Science of Light, Staudtstra\ss e 2, D-91058 Erlangen, Germany}
\affiliation{Department of Physics, Friedrich-Alexander Universit\"at Erlangen-N\"urnberg (FAU), Staudtstra{\ss}e 7,  D-91058 Erlangen, Germany}
\author{Susanne F. Yelin}
\affiliation{Department of Physics, Harvard University, Cambridge, Massachusetts 02138, USA}
\author{Claudiu Genes}
\email{claudiu.genes@physik.tu-darmstadt.de}
\affiliation{TU Darmstadt, Institute for Applied Physics, Hochschulstra\ss e 4A, D-64289 Darmstadt, Germany}
\affiliation{Max Planck Institute for the Science of Light, Staudtstra\ss e 2, D-91058 Erlangen, Germany}

\begin{abstract}
We solve Dicke superradiance with two or more competing collective decay channels of tunable rates using a symbolic quantum-trajectory construction. The method yields closed time-domain populations and observables as finite combinations of exponential terms for arbitrary numbers of emitters and arbitrary decay rates. For two channels, the behavior of the stationary ground-state distribution resembles a first-order phase transition at the point where the channel-rate ratio is equal to unity. For balanced $d$-channel decay, we obtain scaling laws for the superradiant peak time and intensity. These results unify and extend single-channel Dicke dynamics to multilevel emitters and provide a compact tool for cavity and waveguide experiments, where permutation-symmetric reservoirs engineer multiple collective decay paths.
\end{abstract}

\maketitle


Collective light–matter coupling is a cornerstone of quantum optics, with Dicke’s seminal prediction of superradiance standing as its most striking manifestation~\cite{dicke1954}. When an ensemble of $N$ identical two-level emitters is fully inverted, the sample releases a {superradiant burst} whose peak intensity scales quadratically, $I_{\max}\propto N^2$, and occurs at $t_{\mathrm{peak}}\simeq \ln N/(\gamma N)$, where $\gamma$ is the single-atom decay rate~\cite{temnov2005,Scully2009super}. Closed, time-domain solutions of this single-channel problem were obtained in the 1970s~\cite{tsung1977Part1,tsung1977Part2} and by different methods, more recently~\cite{holzinger2025a,holzinger2025b}.

{Real emitters, however, are rarely ideal two-level systems. Many atomic, molecular, and solid-state emitters possess multiple stable ground sublevels. Collective decay of the different optical branches is most faithfully described by Eq.~\eqref{eq:ME_intro} when the emitters couple to a common engineered reservoir, such as a lossy cavity mode or a guided mode of a waveguide. Subwavelength free-space ensembles can also exhibit collective emission, but generally feature additional dipole--dipole interactions and short-range coherent photon exchange that are not included in the permutation-symmetric dissipative model considered here. In the collective-reservoir setting, every branch $|e\rangle\rightarrow|g_\alpha\rangle$ can radiate collectively, yielding a multinomial web of decay paths.} Analytical results for such \emph{multichannel} superradiance exist only in narrow limits—one dominant channel, in the limit of large $N$, or full numerical diagonalization of the Liouvillian superoperator for a few emitters~\cite{sutherland2017superradiance,Orioli2022emergent,masson2024dicke,mok2025,sivan2025multilevel}. A general closed-form solution has been missing.

\begin{figure}[ht!]
  \centering
  \includegraphics[width=1\linewidth]{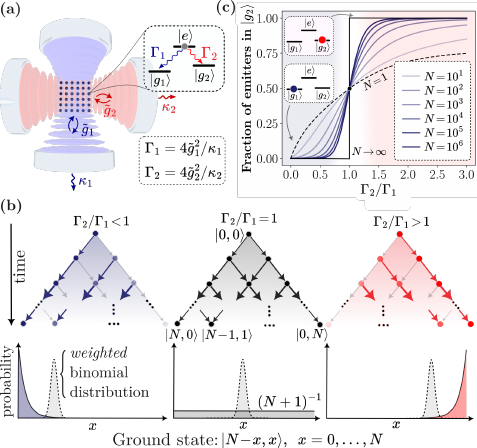}
  \caption{(a) Two-channel Dicke superradiance engineered by two lossy cavity modes acting as permutation-symmetric reservoirs. Each mode couples resonantly to a transition of $N$ $\Lambda$-type emitters with strengths $\tilde{g}_\alpha$ and cavity decays $\kappa_\alpha$, giving collective decay rates $\Gamma_\alpha=4\tilde{g}_\alpha^2/\kappa_\alpha$ in the bad-cavity limit $\kappa_\alpha\gg \tilde{g}_\alpha$. (b) Competition within the ground-state manifold spanned by $\{|g_1\rangle,|g_2\rangle\}$, increasingly sensitive to the decay-rate ratio $r\!=\!\Gamma_2/\Gamma_1$ with growing $N$. Our symbolic trajectory method yields the full time evolution from the inverted state, through the burst, to the \steady{} state distribution of Eq.~\eqref{eq:Multinomial}, in contrast to the weighted binomial distribution for independently decaying emitters. For independent decay $|N\!-\!x,x\rangle$ denotes product states $\ket{g_1}^{\otimes(N-x)}\!\otimes\!\ket{g_2}^{\otimes x}$. (c) In the limit $N\!\rightarrow\!\infty$ the fraction of emitters in $|g_2\rangle$ exhibits a sharp change around $r\!=\!1$.}
  \label{fig:fig1}
\end{figure}

We provide such a solution for $N$ identical, initially fully excited emitters with one excited state $\ket{e}$ and a $d$-fold ground manifold $\{\ket{g_1},\dots,\ket{g_d}\}$. In the absence of driving, the density matrix obeys the permutation-symmetric Lindblad equation
\begin{equation}
  \dot{\hat{\rho}}=\sum_{\alpha=1}^{d}\Gamma_\alpha\,
           \mathcal D[\hat S_\alpha]\,\hat{\rho}, \qquad
  \hat S_\alpha=\sum_{j=1}^{N}\ket{g_\alpha^{(j)}}\!\bra{e^{(j)}},
  \label{eq:ME_intro}
\end{equation}
where $\mathcal D[\hat C]\hat{\rho}=\hat{C}\hat{\rho} \hat{C}^{\dagger}-\tfrac12\{\hat{C}^{\dagger}\hat{C},\hat{\rho}\}$. Because the collective jump operators $\hat{S}_{\alpha}$ commute with
particle permutations and we start in the fully symmetric state
$|e\rangle^{\otimes N}$, the dynamics remain in the symmetric subspace
where $\hat{S}_{\alpha}$ are the lowering operators for $SU(d+1)$ with
known matrix elements. The Hilbert-space dimension grows only
polynomially with $N$, as $\binom{N+d}{d}$.

{
Permutation symmetry has been exploited more generally to reduce the
numerical complexity of open many-body dynamics, including
permutation-invariant master-equation approaches~\cite{Chase2008,Baragiola2010,Xu2013} and quantum-trajectory
formulations in symmetry-reduced Dicke spaces~\cite{Zhang2018,Lloyd2026}. The additional simplification exploited
here is that the present undriven collective-decay dynamics form a
directed, acyclic cascade in excitation number, while the no-jump
evolution is diagonal in the symmetric Dicke basis. The ordered
jump-time integrals can therefore be evaluated analytically as finite
convolutions, or equivalently as rational functions in Laplace space.
Thus, permutation symmetry provides the reduced state space, whereas
the acyclic trajectory structure enables the closed analytical
solution.
}

We derive a closed time-domain expression for $\hat{\rho}(t)$ valid
for arbitrary $N$ and arbitrary decay rates $\{\Gamma_\alpha\}$.
Our approach is built on a symbolic quantum-trajectory
(quantum-jump) construction~\cite{dalibard1992,Carmichael1993,Plenio1998},
adapted to the multichannel case. All observables, populations,
intensities, and higher-order correlations can then be obtained
analytically from the resulting finite trajectory expressions.

Previous work on multilevel superradiant dynamics has shown, for example, cavity-mediated decay of multilevel atoms into permutation-symmetric entangled dark states~\cite{Orioli2022emergent}. Mok \emph{et al.} analyzed many-body decay with competing branches and proved a “winner-takes-all’’ ground-state selection from a hydrodynamic continuum description~\cite{mok2025}. In contrast, we obtain exact closed solutions of the $d$-channel Dicke master equation via a symbolic trajectory construction, giving full transient dynamics and exact \steady{} state distributions for any $N$, which are not available in Refs.~\cite{Orioli2022emergent,mok2025}. Note that the \steady{} state here refers to the $t\to\infty$ limit of the undriven Lindblad dynamics, which depends on the initial state.

Figure~\ref{fig:fig1}(a) illustrates a possible implementation of multichannel Dicke superradiance via lossy cavity modes resonant with the emitters' decay branches. These modes act as permutation-symmetric reservoirs that funnel all radiation into collective channels~\cite{park2022cavity}. For resonant cavity modes with decay rates $\kappa_\alpha$ coupled with strengths $\tilde{g}_\alpha$ to transitions
$\ket{e}\!\to\!\ket{g_\alpha}$, adiabatic elimination in the bad-cavity regime
$\kappa_\alpha \gg \tilde{g}_\alpha$ yields a purely dissipative Dicke master equation with collective jump operators
$\hat S_\alpha$ and cavity-tuned rates $\Gamma_\alpha=4\tilde{g}_\alpha^{2}/\kappa_\alpha$, without collective Lamb shift (it vanishes at resonance; see Supplemental Material~\cite{SM}). Two orthogonal polarizations of a single cavity, or two crossed cavities, generate a two-channel scenario with independently tunable rates $\Gamma_{1,2}$ and adjusting $\tilde{g}_\alpha$ (or $\kappa_\alpha$) sets the ratio
$r=\Gamma_2/\Gamma_1$ that controls the \steady{} state ground-state distribution in Fig.~\ref{fig:fig1}(b,c). For large $N$ this leads to a sharp transition around the balanced decay point $r\!=\!1$, where the ground-state distribution is flat, in contrast to the binomial distribution for independent decay. The cavity setup is a far more faithful implementation of the model Eq.~\eqref{eq:ME_intro} than emitters in a small volume, where the vectorial character of the dipolar interactions and the diverging short-range unitary photon exchange complicate the dynamics.

Dicke superradiance in the bad-cavity regime has been realized with cold atoms (steady-state Raman superradiant lasers and ultranarrow Sr transitions) and in circuit QED with artificial atoms in fast-decaying cavities~\cite{Bohnet2012,norcia2016prx,norcia2016superradiance,norcia2018prx,mlynek2014}. Multichannel Dicke superradiance can also be realized in one-dimensional nanophotonic waveguide QED platforms~\cite{Brekenfeld2020crossed,Goban2015PRL,Solano2017superradiance,sheremet2023waveguide,Brehm2021NPJQM,Liedl2024PRX}, where guided modes provide permutation-symmetric reservoirs. In the Supplemental Material~\cite{SM} we show the adiabatic elimination of multiple cavity modes and its validity in the bad-cavity limit.

\prlsection{Introducing the symbolic quantum trajectory method}
We first introduce the symbolic quantum-trajectory method for a single jump operator with rate $\Gamma$, corresponding to single-channel Dicke superradiance, for which analytical solutions are known~\cite{holzinger2025a,holzinger2025b}. The method yields an exact time-domain form of $\hat{\rho}(t)$ obtained from quantum trajectories. The stochastic unraveling converts the Lindblad equation into an ensemble of pure-state trajectories~\cite{dalibard1992,few_atoms_charmichael,charmichael_2}; here we summarize the elements needed.

\begin{figure}[t]
  \centering
  \includegraphics[width=0.98\linewidth]{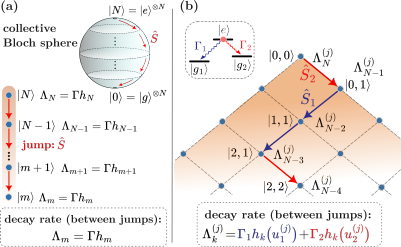}
  \caption{(a) \textit{Single-channel Dicke superradiance.} Decay cascade from the fully inverted state: non-Hermitian time evolution with rates $\Lambda_m$ and successive jumps ($\hat{S}$) generate states $|m\rangle$ [Eq.~\eqref{eq:trajSingle}]. The dynamics reside on the surface of the collective Bloch sphere. (b) \textit{Two-channel Dicke superradiance.} An inverted ensemble of $N$ three-level systems with a twofold ground manifold decays through two collective channels at rates $\Gamma_1$, $\Gamma_2$. Starting from $|0,0\rangle$, a trajectory $(j)$ with non-Hermitian evolution (rate $\Lambda_{k}^{(j)}$) and jumps ($\hat{S}_1$, $\hat{S}_2$) through states $|u_1^{(j)},u_2^{(j)}\rangle$ ends in $|n_1,n_2\rangle$ [Eq.~\eqref{eq:two-channel}]. The ground state is a distribution over $n_1+n_2=N$.}
  \label{fig:level}
  \vspace{-1em}
\end{figure}
In the quantum-jump picture, a trajectory is a sequence of nonunitary evolutions interrupted by instantaneous jumps. Let $\hat{S}$ be the collective collapse operator with rate $\Gamma$ [Eq.~\eqref{eq:ME_intro} and Fig.~\ref{fig:level}(a)]. Between jumps the system evolves under
$\hat U_{t,t'} = \exp[{-i\hat H_{\mathrm{eff}}\,(t-t')}]$ with effective non-Hermitian Hamiltonian $\hat H_{\mathrm{eff}} = - i\frac{\Gamma}{2} \hat{S}^\dagger \hat{S}$. Any additional diagonal term $\hat H_{0}$ (containing the level energies) commuting with $\hat S^{\dagger}\hat S$ would not affect the population dynamics considered here. Let $\hat{S}$ act on Dicke states as $\hat{S}\ket{m} = \sqrt{h_m}\ket{m-1}$ with $h_m = m(N+1-m)$, where $\ket{m}$ denotes the symmetric Dicke state with $m$ excitations. The density operator is diagonal in this basis, $\hat{\rho}(t)=\sum_{m=0}^N p_m(t) |m\rangle \langle m |$.

Consider a trajectory at time $t$ starting from $\ket{N}$ at $t_0\!=\! 0$ with exactly $q$ jumps at times $t_0 < t_1 < \dots < t_q < t$. After $q$ jumps the state is $\ket{m}$ with $m=N-q$ excitations:
\begin{align}
\ket{\psi_q(t;\boldsymbol{t}_q)}
= \hat{U}_{t,t_q}\hat{S}
\cdots
\hat{S}\,\hat{U}_{t_2,t_1}\hat{S}\,\hat{U}_{t_1,t_0}\ket{N},
\label{eq:trajSingle}
\end{align}
where $\boldsymbol{t}_q=\{t_1,...,t_q \}$ is the set of jump times. The $q$ actions of $\hat S$ give the prefactor
\begin{align}
\sqrt{ \Gamma^{N-m} h_N\cdots h_{m+1}} \!=\! \sqrt{\frac{\Gamma^{N-m}N!(N\!-\!m)! }{m!}} \!=\! \sqrt{\mathcal{C}_m},
\end{align}
where each collapse operator includes a factor $\sqrt{\Gamma}$~\cite{dalibard1992}. The $q\!+\!1$ non-Hermitian evolution operators yield a product of exponentials with decay rates of the visited states: $e^{-\frac{\Gamma}{2} h_{N-q} (t-t_q)}\cdots e^{-\frac{\Gamma}{2} h_N t_1 }$.

The density operator is obtained by summing \textit{all} trajectories and integrating over ordered jump times:
\begin{equation}
\hat\rho(t) \!=\!
\sum_{q=0}^{N}\;
 \int\displaylimits_{0<t_1<\dots<t_q<t}\!\!\!
dt_1\!\dots dt_q\ \!
\ket{\psi_q(t;\boldsymbol{t}_q)}\!
\bra{\psi_q(t;\boldsymbol{t}_q)}.
\label{eq:rho_sum}
\end{equation}
The population $p_m(t)$ is therefore
\begin{equation}
p_m(t)
= \mathcal{C}_m
\!\!\!
\int\displaylimits_{0<t_1<\dots<t_q<t}
e^{-\Lambda_m (t - t_q)} \cdots e^{-\Lambda_N t_1 }
\,dt_1\cdots dt_q,
\end{equation}
where $q = N-m$ and with rates $\Lambda_m = \Gamma h_m$. More formally and compactly, we can rewrite the solution as a sequence of nested convolutions (using $(f*g)(t)=\int_0^t f(t-\tau)g(\tau)d\tau$ as the definition for convolutions):
\begin{equation} \label{eq:ds_convolution}
p_m(t) =
\mathcal{C}_m \,
\left(
  e^{-\Lambda_N t}
   *  \cdots
   *  e^{-\Lambda_m t}
\right).
\end{equation}
Each convolution adds an exponential with the decay rate of the respective state. For a target state $m$ the sequence contains $N - m + 1$ terms [Fig.~\ref{fig:level}(a)]. For states below half inversion, degeneracy $\Lambda_m=\Lambda_{N+1-m}$ leads to repeated terms, and by commutativity, one finds $e^{-\Lambda_m t} * e^{-\Lambda_{N+1-m}t}= t e^{-\Lambda_m t}$. Alternatively, one may work in Laplace space:
\begin{equation}
\tilde{p}_{m}(s) = \mathcal{C}_{m}\prod_{k=m}^{N} \frac{1}{(s + \Lambda_k)}, \quad
p_m(t) =\frac{1}{2\pi i}\oint_{\mathcal C}
{\tilde{p}_m(s)} e^{s t}\,ds,
\label{eq:laplace}
\end{equation}
where $\mathcal C$ encircles all poles $s=-\Lambda_k$ counter-clockwise. For $m$ below the equator, $\Lambda_m=\Lambda_{N+1-m}$ yields double poles~\cite{holzinger2025a}. If $\{\Lambda_\alpha\} \!=\! \{\Lambda_k:\;k=m,\dots,N\}$ is the set of distinct rates with multiplicities $\nu_\alpha\in\{1,2\}$, one can equivalently replace $\tilde{p}_m(s)$ by $\mathcal Q_m(s)=\mathcal{C}_m\prod_{\alpha}(s+\Lambda_\alpha)^{-\nu_\alpha}$~\cite{holzinger2025b,holzinger2025a}.

For instance, the standard signature of superradiance, namely the emitted power, $\mathcal{I} (t)=\text{Tr}\left[\Gamma\hat S^\dagger \hat S\hat \rho(t)\right]$ (in units of $\hbar\omega$), becomes
\begin{equation}
    \mathcal{I} (t)=\sum_{m=0}^N \Lambda_m p_m(t)=\Gamma\sum_{m=0}^N m(N+1-m)\; p_m(t).
\end{equation}
We illustrate the standard peak of
superradiance ($d=1$) and its extension to multiple decay
channels in Fig.~\ref{fig3}.

\prlsection{Two-channel Dicke superradiance}
We now generalize to Dicke superradiance with two competing channels $\ket{e} \to \ket{g_1}$ and $\ket{e} \to \ket{g_2}$ with decay rates $\Gamma_1$ and $\Gamma_2$ and jump operators $\hat{S}_1$ and $\hat{S}_2$ defined in Eq.~\eqref{eq:ME_intro}.

Since $\hat{S}_{1,2}$ commute with all particle permutations and the initial state $\ket{e}^{\otimes N}$ is symmetric, the dynamics reside in the symmetric subspace of
$\bigl(\mathbb{C}^{3}\bigr)^{\otimes N}$.
As in the single-channel case we track states with $m$ excitations, starting from $m=N$. We label each symmetric Dicke state by the occupation numbers $(n_1,n_2)$ of ground sublevels $\ket{g_1}, \ket{g_2}$, so that after $q$ jumps we have $m=N-n_1-n_2=N-q$ excited emitters [Fig.~\ref{fig:level}(b)]. We define
\begin{align}
\ket{n_1,n_2}
  =
  \mathcal{N}\,
  \sum_{\pi\in S_N}
  P_\pi\bigl(
  \ket{e}^{\otimes m}
  \otimes \ket{g_1}^{\otimes n_1}
  \otimes \ket{g_2}^{\otimes n_2}
  \bigr),
\label{eq:Multinomial}
\end{align}
where $P_\pi$ permutes emitters and $S_N$ is the symmetric group. The normalization $\mathcal{N}=\sqrt{m!\,n_1!\,n_2!/N!}$ ensures $\braket{n_1',n_2'|n_1,n_2}
=\delta_{n_1 n_1'}\delta_{n_2 n_2'}$.

We let the excitation number $m$ run from $N$ down to $0$ and introduce counters $u_1\in[0,n_1]$ and $u_2\in[0,n_2]$. At each level $m$ they obey $u_1+u_2=N-m$, starting at $(0,0)$ and ending at $(n_1,n_2)$ after $q$ jumps. Introducing $h_k(x)=k(x+1)$ simplifies the collective action:
\begin{equation}
\begin{aligned}
\hat{S}_1\ket{u_1, u_2}
   &= \sqrt{h_k(u_1)}\;\ket{u_1\!+\!1,\,u_2},\\[2pt]
\hat{S}_2\ket{u_1, u_2}
   &= \sqrt{h_k(u_2)}\;\ket{u_1,\,u_2\!+\!1},
\end{aligned}
\end{equation}
with $k=N-u_1-u_2$. We follow the evolution on each decay path from $\ket{0,0}$ to $\ket{n_1,n_2}$. A path $j$ is specified by a string $\alpha=(\alpha_N,\cdots,\alpha_{m+1})$ of length $q=N-m$, where $\alpha_k=1$ ($0$) denotes a jump through $\hat S_1$ ($\hat S_2$). The number of distinct paths is
$n_\text{paths}={(n_1+n_2)!}/({n_1!\,n_2!})$.

Along any path $j$, the collapse operators generate a product containing $n_1$ factors from channel 1 and $n_2$ from channel 2. This gives a product $N\cdots (m+1)=N!/m!$ multiplied with a product $n_1! n_2!$, which we collect as
$\mathcal{C}_{n_1,n_2}=\frac{N!\,n_1!\,n_2!\Gamma_1^{n_1}\Gamma_2^{n_2}}{m!}$. The non-Hermitian evolution picks up the total rates out of the states reached along path $j$,
\begin{equation}\label{eq:rate_eq_main}
\begin{split}
\Lambda_k^{(j)}
&= \Gamma_1\, h_k\!\big(u_1^{(j)}\big) + \Gamma_2\, h_k\!\big(u_2^{(j)}\big) \\[2pt]
&= k\!\left[\Gamma_1\!\left(u_1^{(j)} + 1\right) + \Gamma_2\!\left(u_2^{(j)} + 1\right)\right],
\end{split}
\end{equation}
as illustrated in Fig.~\ref{fig:level}(b). Summing over all paths gives
\begin{widetext}
\vspace{-1em}
\begin{align}
\label{eq:two-channel}
p_{n_1,n_2}(t) =
\mathcal{C}_{n_1,n_2}\;
\left(
e^{-\Lambda_{N} t}
*
\left[
\sum_{j=1}^{n_{\text{paths}}}
  e^{-\Lambda^{(j)}_{N-1} t}
  * \cdots *
  e^{-\Lambda^{(j)}_{m+1} t}
\right]
*
e^{-\Lambda_m t}
\right).
\end{align}
\vspace{-1em}
\end{widetext}
The endpoints $\ket{0,0}$ and $\ket{n_1,n_2}$ are common to all trajectories, so $\Lambda_{N}=N(\Gamma_1+\Gamma_2)$ and $\Lambda_{m}=m\left[\Gamma_1 (n_1+1)+\Gamma_2(n_2+1)\right]$ are path-independent. The Laplace transform $\tilde{p}_{n_1,n_2}(s)$ is obtained from Eq.~\eqref{eq:two-channel} analogously to Eq.~\eqref{eq:laplace}, now with a sum over paths:
{
\begin{equation}
    \tilde{p}_{n_1,n_2}(s)=\mathcal{C}_{n_1,n_2} \sum_{j=1}^{n_\mathrm{paths}} \prod_{k=m}^N \frac{1}{s+\Lambda^{(j)}_k},
\end{equation}
with $m=N-n_1-n_2$.}
{
For the stationary ground-manifold state $m=0$, the final segment has $\Lambda_0^{(j)}=0$ for every path. Hence the Laplace transform contains one factor $1/s$, which leads to the steady state
\begin{align}
\label{eq:ground-state-steady}
    {p}^{(ss)}_{n_1,n_2} &= \lim_{s\rightarrow0}s~ \tilde{p}_{n_1,n_2}(s)= \mathcal{C}_{n_1,n_2} \sum_{j=1}^{n_\mathrm{paths}} \prod_{k=1}^N \frac{1}{\Lambda^{(j)}_k} .
\end{align}
Hence, the nontrivial stationary ground-manifold distribution follows directly from the same Laplace-space trajectory solution as the transient dynamics, without requiring an inverse Laplace transform or numerical long-time propagation.
}

{As a simple example, let us consider $\Gamma_1=\Gamma_2={\Gamma}/2$. The rates $\Lambda_k^{(j)}\!=\!{\Gamma}k(N\!-\!k\!+\!2)/2$ become completely independent of the path, and one obtains $\prod_{k=1}^N \Lambda_k^{(j)}= ({\Gamma}/{2})^N N!(N+1)!$, leading to the flat ground state distribution for balanced decay,
\begin{equation}
  p_{n_1,n_2}^{(ss)} = \frac{1}{N+1} .
\end{equation}}
The generalization to $d>2$ collective decay channels, and analytic expressions for the \steady{} state $p_{n_1,...,n_d}^{(ss)}$, are given in the Supplemental Material~\cite{SM}.

\prlsection{Superradiant transient and \steady{} state}
In Fig.~\ref{fig3}(a) we show Dicke superradiance for $d$ collective decay channels with balanced rates, $\Gamma_1=\cdots=\Gamma_d=\Gamma/d$. {For balanced decay, all trajectories that reach a fixed excitation number $m$ experience the same total non-Hermitian decay rate, independent of how the previous jumps were distributed among the $d$ ground-state channels. The multichannel dynamics therefore collapse to an effective one-dimensional cascade in $m$.} The emitted intensity $\mathcal I_{\mathrm{tot}}(t)=(\Gamma/d)\sum_{\alpha} \langle \hat{S}_\alpha^\dagger \hat{S}_\alpha \rangle$ (in units of $\hbar\omega$) is plotted versus time for increasing $d$. As $d$ grows, the burst weakens and shifts to later times. In Fig.~\ref{fig3}(b), we plot the total emitted peak intensity and delay time~\cite{gross_haroche} as functions of $d$. We find
\begin{equation} \label{eq:theory}
    \frac{\mathcal I_{\text{tot}}^{\max}}{\Gamma}
    \approx \frac{(N+d-1)^2}{4d+1},
    \qquad
    \Gamma t_{\text{peak}}
    \approx
    \frac{d\,\ln\!\bigl(\tfrac{N}{d}\bigr)}{N+d}.
\end{equation}
The peak-intensity expression in Eq.~\eqref{eq:theory} is obtained by fitting the peak values extracted from the exact analytical solution, whereas the peak time is calculated in the Supplemental Material~\cite{SM} from the expected time to reach the state with $m=(N+d)/2$. Equation~\eqref{eq:theory} reproduces the standard $d=1$ scaling and captures the $1/d$ suppression of the peak intensity for $d$ equal decay channels. The delayed and weakened superradiant burst follows directly from the loss rate $\tfrac{\Gamma}{d}m(N-m+d)$ of a state with $m$ excitations, which yields a parabolic scaling $\sim N^2/d$ around $m=(N+d)/2$ and therefore suppresses the peak as $d$ increases.

For single-channel Dicke superradiance the \steady{} state is trivial, $|\psi(t\rightarrow\infty)\rangle=|g\rangle^{\otimes N}$. For $d=2$, the \steady{} state is a distribution over ground states $|n_1,n_2\rangle$ with $n_1\!+\!n_2\!=\!N$ [Fig.~\ref{fig:fig1}(b) and Eq.~\eqref{eq:Multinomial}]. We provide a closed-form expression for $p^{(ss)}_{n_1,n_2}$ valid for any $N$ and $\Gamma_{1,2}$ in the Supplemental Material~\cite{SM}. Writing these probabilities as $p^{(ss)}_{N-x,x}$ with $x\!= \! 0,\dots,N$, we define the \steady{} state population fraction in $|g_2\rangle$ as
$\bar{n}_2 = \frac{1}{N}\sum_{x=0}^N x \, p^{(ss)}_{N-x,x}$,
with control parameter $r=\Gamma_2/\Gamma_1$, see Fig.~\ref{fig3}(c).

\begin{figure}[ht!]
  \centering
  \includegraphics[width=0.95\linewidth]{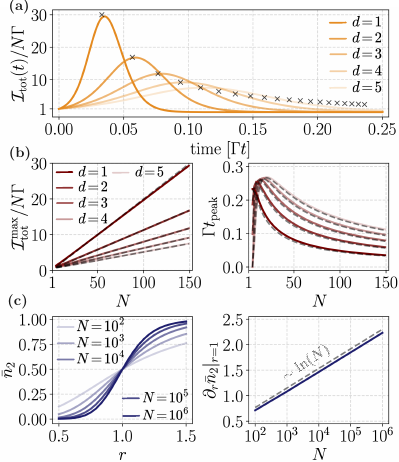}
    \vspace{-0.5em}
  \caption{(a) Emitted intensity versus time for increasing numbers of collective decay channels $d$ with balanced rates $\Gamma_1=\cdots=\Gamma_d=\Gamma/d$ for $N=150$. As $d$ grows, the burst weakens and is delayed. Crosses indicate analytic predictions for $(t_{\mathrm{peak}},\mathcal I_{\mathrm{tot}}^{\max})$ from Eq.~\eqref{eq:theory}. (b) Total emitted peak intensity and peak time versus $d$ from the analytic trajectory solution (Supplemental Material~\cite{SM}). Dashed lines show the analytic scalings in Eq.~\eqref{eq:theory}. (c) Final population fraction $\bar{n}_2$ in $|g_2\rangle$ versus decay-rate ratio $r$. The order parameter changes sharply near $r\!=\!1$, with analytic slope $\partial_r \bar{n}_2|_{r=1}\approx \ln (N)/6$ in Eq.~\eqref{eq:slope}.}
  \label{fig3}
  \vspace{-1.5em}
\end{figure}
\prlsection{First-order-like dissipative transition}
The \steady{} state distribution $p^{(ss)}_{N-x,x}$ reached from the fully excited state resembles a first-order phase transition~\cite{Carmichael2015breakdown,Beaulieu2025DPTKerr}. What we call a phase transition here is a non-analyticity of the order parameter $\bar n_2(N,r)$ at $r=1$ as $N\to\infty$, not necessarily accompanied by a spontaneous symmetry breaking as in equilibrium systems. The order parameter is $\bar{n}_2(N,r)$ and the control parameter is the decay-rate ratio $r$. At $r=1$ symmetry fixes $\bar{n}_2=1/2$ for any $N$, while for $r\neq 1$ the distribution is strongly biased. In the limit $N\to\infty$, $\bar{n}_2(N,r)$ sharpens into a Heaviside step function, as in a first-order transition. The mechanism is simple: decay into a particular ground state $\ket{g_\alpha}$ enhances the probability that subsequent decays populate the same state, so an infinitesimal rate imbalance becomes amplified in the ground-state distribution.
Quantitatively, the susceptibility at the balanced point follows in closed form as (see Supplemental Material~\cite{SM})
\begin{equation}\label{eq:slope}
\left.\partial_r \bar n_2(N,r)\right|_{r=1}
= \frac{\ln N}{6}+{O}\!\left(1\right).
\end{equation}

However, the thermodynamic analogy breaks down here: Because the number of accessible states grows linearly with $N$, the normalized distribution $p_{N-x,x}^{(ss)}$ appears locally flat as $N\to\infty$, corresponding to an effective $T=\infty$. The stationary behavior map onto a two-color Pólya urn with unequal reinforcement rates~\cite{janson2004functional}, which provides a natural stochastic-process description of the reinforcement mechanism underlying the logarithmically increasing susceptibility.

{We note that generic symmetry-breaking perturbations may round the large-$N$ nonanalytic behavior, so the present solution provides the permutation-symmetric limiting case for assessing the robustness of this transition.}

\prlsection{Discussion and conclusions}
We introduced a symbolic quantum-trajectory solution of Dicke superradiance with $d$ competing collective decay channels. The method yields closed time-domain expressions for symmetric-state populations and observables as finite closed combinations of exponential terms with rates set by $N$ and $\{\Gamma_\alpha\}$. For balanced multichannel emission we derived compact analytic scalings for the superradiant delay time and peak emission rate, Eq.~\eqref{eq:theory}, extending Dicke’s $d{=}1$ results to multilevel emitters. For two channels we found a sharp ground-manifold selection controlled by $r=\Gamma_2/\Gamma_1$, with logarithmically diverging susceptibility at the balanced point, Eq.~\eqref{eq:slope}, and a Heaviside-like order parameter in the limit $N\to\infty$.

Experimentally, crossed or polarization-resolved cavities enable tunable multichannel superradiance with adjustable $r$~\cite{Brekenfeld2020crossed}, while bad-cavity superradiance with cold atoms has been realized on narrow optical transitions~\cite{Bohnet2012,norcia2016prx}. In nanophotonic waveguide QED, guided modes provide permutation-symmetric reservoirs and polarization/multimode engineering provides tunable channel rates~\cite{Goban2015PRL,Solano2017superradiance,sheremet2023waveguide}. These settings allow initialization near $\ket{e}^{\otimes N}$ (or rapid pumping), control of $r$, and time-resolved detection of both the multichannel burst and the final ground-manifold distribution. The resulting “winner-takes-all”~\cite{mok2025} ground-state selection resembles a dissipative many-body phase transition: the \steady{} state of the Liouvillian acquires a nonanalytic dependence on the ratio of cooperative decay rates, with a susceptibility that diverges logarithmically with $N$. This connects multichannel Dicke superradiance to experimentally studied nonequilibrium critical phenomena in driven-dissipative platforms~\cite{Fink2017NatPhys,Carmichael2015breakdown,minganti2018spectral,Sieberer2016RPP}, while keeping the control knob minimal (a ratio of collective decay rates).

{Permutation-invariant local decoherence or pumping can instead be incorporated within weak-permutation-symmetry formulations~\cite{Chase2008,Baragiola2010,Xu2013,Lloyd2026}. In the trajectory picture, incoherent pumping introduces
loops into the otherwise acyclic jump graph and therefore requires a resummation or recursive treatment, while coherent driving generally replaces the scalar no-jump evolution terms by matrix-valued terms within the symmetric manifold, as encountered
in fully collective driven models~\cite{Reilly2026}. Generic emitter-dependent disorder or inhomogeneous couplings, by contrast, break the symmetry underlying the present exact solution. The model solved here therefore provides both an analytically tractable limit and a reference point for studying how such perturbations modify multichannel collective decay.}

\vspace*{0.1cm}
\noindent \textbf{Acknowledgments -} We acknowledge financial support from the Max Planck Society and the Deutsche Forschungsgemeinschaft (DFG, German Research Foundation) -- Project-ID 429529648 -- TRR 306 QuCoLiMa
(``Quantum Cooperativity of Light and Matter''). SFY acknowledges the NSF via PHY-2207972. This research was funded in whole or in part by the Austrian Science Fund (FWF) [10.55776/COE1].

\title{Supplemental Material: Symbolic Quantum-Trajectory Method for Multichannel Dicke Superradiance}

\author{Raphael Holzinger}
\email{raphael.holzinger@uibk.ac.at}
\affiliation{Department of Physics, Harvard University, Cambridge, Massachusetts 02138, USA}
\author{Nico S. Bassler}
\affiliation{TU Darmstadt, Institute for Applied Physics, Hochschulstra\ss e 4A, D-64289 Darmstadt, Germany}
\author{Julian Lyne}
\affiliation{Max Planck Institute for the Science of Light, Staudtstra\ss e 2, D-91058 Erlangen, Germany}
\affiliation{Department of Physics, Friedrich-Alexander Universit\"at Erlangen-N\"urnberg (FAU), Staudtstra{\ss}e 7,  D-91058 Erlangen, Germany}
\author{Susanne F. Yelin}
\affiliation{Department of Physics, Harvard University, Cambridge, Massachusetts 02138, USA}
\author{Claudiu Genes}
\email{claudiu.genes@physik.tu-darmstadt.de}
\affiliation{TU Darmstadt, Institute for Applied Physics, Hochschulstra\ss e 4A, D-64289 Darmstadt, Germany}
\affiliation{Max Planck Institute for the Science of Light, Staudtstra\ss e 2, D-91058 Erlangen, Germany}

\maketitle 

\clearpage
\pagebreak
\onecolumngrid

\setcounter{section}{0}
\setcounter{figure}{0}
\setcounter{table}{0}\renewcommand{\thetable}{A\arabic{table}}
\setcounter{equation}{0}\renewcommand{\theequation}{A\arabic{equation}}

\vspace{-3em}


\section{Adiabatic elimination of multiple cavity modes}

Suppose we couple a system with $d$ ground states as introduced in the main text with operators $\hat S_\alpha$ to $d$ bad cavity modes $\hat a_\alpha$ with couplings $g_\alpha$ and decay rates $\kappa_\alpha$, where each cavity is on-resonance with the transition, then the Hamiltonian for the system is
\[
H_I/\hbar =\sum_{\alpha=1}^d g_\alpha \big(\hat a_\alpha^\dagger \hat S_\alpha^{\vphantom\dagger}+\hat a_\alpha^{\vphantom\dagger}\hat S_\alpha^\dagger \big),
\]
assuming rotating wave approximations for all transitions. Then the Master equation for the system is given by
\[
\dot{\rho}=-\frac{\ii}{\hbar}\commutator{H_I}{\rho}+\sum_{\alpha=1}^d \kappa_\alpha D[\hat a_\alpha].
\]

A standard elimination of the cavity modes $\hat a_\alpha$ under Born-Markov approximation, i.e. $g_\alpha\ll\kappa_\alpha$, lets us find the effective system dynamics for the atomic transitions as
\begin{equation}
    \dot\rho=\sum_{\alpha=1}^{d}\Gamma_\alpha D[S_\alpha],
\end{equation}
with the decay rate (where $\expval{}_B$ indicates the evaluation of the correlation under the assumption of free bath evolution)
\begin{equation}
    \Gamma_\alpha=g_\alpha^2\Re\int_{-\infty}^{\infty}\text{d}t \expval{\hat a_\alpha(t)\hat a_\alpha^\dagger(0)}_B=\frac{4g_\alpha^2}{\kappa_\alpha},
\end{equation}
as defined in the main text.

\section{Solving Dicke superradiance for more than two channels}
\label{App:Multichannel}

Let $\vec n=(n_{1},\dots,n_{d})$ denote the number of decay events in
each channel, with $m$ remaining excitations and the conservation condition
\[
m=N-\sum_{\alpha=1}^{d}n_\alpha.
\]
A quantum trajectory that realizes this final configuration is specified
by an ordered list of jumps
\[
\boldsymbol{\alpha}^{(j)}
=
\bigl(\alpha^{(j)}_1,\dots,\alpha^{(j)}_{q}\bigr),
\qquad q=N-m,
\]
where $\alpha^{(j)}_\ell\in\{1,\dots,d\}$ specifies the decay channel of
the $\ell$-th jump. Because the sequence matters, we attach a trajectory
index $j$ to every path-dependent quantity.

After $N-k$ jumps, the system occupies the Dicke level $k$.
Define the counters $u_{k}^{(\alpha,j)}$ as the number of times branch
$\alpha$ has been used \emph{before} reaching level $k$ in trajectory $j$.
The total non-Hermitian decay rate for the $(j,k)$-th segment is
\begin{equation}
\Lambda^{(j)}_{k}
=
k\sum_{\alpha=1}^{d}
\Gamma_\alpha\left(u_k^{(\alpha,j)}+1\right).
\label{eq:Lambda_general}
\end{equation}
The two endpoints are the same for every trajectory,
\begin{equation}
\Lambda^{(j)}_{N}
=
N\sum_{\alpha=1}^{d}\Gamma_\alpha,
\end{equation}
and
\begin{equation}
\Lambda^{(j)}_{m}
=
m\sum_{\alpha=1}^{d}\Gamma_\alpha(n_\alpha+1).
\end{equation}

Summing over \emph{all} multinomial paths produces the population of a
target state,
\begin{equation}
p_{n_1,\dots,n_d}(t)
=
\mathcal{C}_{n_1,\dots,n_d}
\sum_{j=1}^{n_{\mathrm{paths}}}
\left[
e^{-\Lambda^{(j)}_{N}t}
*
e^{-\Lambda^{(j)}_{N-1}t}
*
\cdots
*
e^{-\Lambda^{(j)}_{m}t}
\right],
\label{eq:pvecn_full}
\end{equation}
with
\begin{equation}
\mathcal{C}_{n_1,\dots,n_d}
=
\Gamma_1^{n_1}\cdots\Gamma_d^{n_d}
\frac{N!\,n_1!\cdots n_d!}{m!}.
\label{eq:C_general}
\end{equation}
The number of distinct paths is the corresponding multinomial coefficient,
\begin{equation}
n_{\mathrm{paths}}
=
\frac{(n_1+\cdots+n_d)!}{n_1!\cdots n_d!}.
\end{equation}

The convolution form in Eq.~\eqref{eq:pvecn_full} becomes particularly
compact in Laplace space. Using
\[
\mathcal{L}\!\left[e^{-\Lambda t}\right](s)
=
\frac{1}{s+\Lambda},
\]
we obtain
\begin{equation}
{
\widetilde p_{n_1,\dots,n_d}(s)
=
\mathcal{C}_{n_1,\dots,n_d}
\sum_{j=1}^{n_{\mathrm{paths}}}
\prod_{k=m}^{N}
\frac{1}{s+\Lambda_k^{(j)}} .
}
\label{eq:laplace_general}
\end{equation}
Thus, each trajectory $j$ contributes a rational function whose poles are
given directly by the non-Hermitian decay rates encountered along that
trajectory.

For completeness, Eq.~\eqref{eq:laplace_general} can also be written over
a common denominator. Let
\[
\mathcal{S}
=
\{\Lambda_1,\ldots,\Lambda_L\}
\]
denote the set of distinct decay rates appearing across all trajectories,
and let $\nu_\ell^{(j)}$ denote the multiplicity with which the rate
$\Lambda_\ell$ occurs in trajectory $j$. Define
\[
\nu_\ell
=
\max_j \nu_\ell^{(j)}.
\]
A common denominator is then
\begin{equation}
Q_{\vec n}(s)
=
\prod_{\ell=1}^{L}
(s+\Lambda_\ell)^{\nu_\ell},
\label{eq:Q_general}
\end{equation}
while the corresponding numerator is
\begin{equation}
P_{\vec n}(s)
=
\mathcal{C}_{\vec n}
\sum_{j=1}^{n_{\mathrm{paths}}}
\prod_{\ell=1}^{L}
(s+\Lambda_\ell)^{\nu_\ell-\nu_\ell^{(j)}}.
\label{eq:P_general}
\end{equation}
Hence
\begin{equation}
\widetilde p_{\vec n}(s)
=
\frac{P_{\vec n}(s)}{Q_{\vec n}(s)},
\qquad
p_{\vec n}(t)
=
\frac{1}{2\pi i}
\oint_{\mathcal C}
ds\,
e^{st}
\frac{P_{\vec n}(s)}{Q_{\vec n}(s)},
\label{eq:contour_general}
\end{equation}
where the contour $\mathcal C$ encloses all poles
$s=-\Lambda_\ell$. In the absence of coinciding rates,
$\nu_\ell=1$ and Eqs.~\eqref{eq:Q_general}--\eqref{eq:P_general}
reduce to products of simple linear factors.

For a stationary ground-manifold state, $m=0$, every trajectory contains
the terminal rate
\[
\Lambda_0^{(j)}=0.
\]
Consequently, each term in Eq.~\eqref{eq:laplace_general} contains one
factor $1/s$, and the final-value theorem directly gives
\begin{equation}
{
p_{n_1,\dots,n_d}^{(\mathrm{ss})}
=
\lim_{s\rightarrow0}
s\,\widetilde p_{n_1,\dots,n_d}(s)
=
\mathcal{C}_{n_1,\dots,n_d}
\sum_{j=1}^{n_{\mathrm{paths}}}
\prod_{k=1}^{N}
\frac{1}{\Lambda_k^{(j)}} ,
}
\qquad
\sum_{\alpha=1}^{d}n_\alpha=N.
\label{eq:ss_general}
\end{equation}
Thus, the nontrivial ground-manifold distribution follows directly from
the same Laplace-space trajectory solution as the transient dynamics:
for each path, one takes the inverse product of all nonzero rates and sums
over all paths terminating in the target state. No inverse Laplace
transformation is required in the stationary limit.

We note that the nested convolutions in
Eq.~\eqref{eq:pvecn_full} can also be evaluated explicitly according to
the general rules
\begin{equation}
e^{-F t} \ast e^{-K t}
=
\begin{cases}
\dfrac{e^{-F t}-e^{-K t}}{K-F}, & F\neq K,\\[6pt]
t\,e^{-F t}, & F=K,
\end{cases}
\end{equation}
and
\begin{equation}
\bigl(t\,e^{-F t}\bigr)\ast e^{-K t}
=
\dfrac{e^{-K t}-e^{-F t}}{(K-F)^2}
-
\dfrac{t\,e^{-F t}}{K-F},
\qquad F\neq K.
\end{equation}
Repeated rates correspond equivalently to higher-order poles in
Eq.~\eqref{eq:contour_general} and lead to polynomial factors multiplying
the corresponding exponential terms.

\vspace{2em}

\section{Balanced decay: $\Gamma_1=\cdots=\Gamma_d=\Gamma/d$}

With identical rates, every trajectory that reaches a fixed Dicke level $k$ experiences the \emph{same} segment rate
\begin{equation}
\Lambda_k=\frac{\Gamma}{d}\;k\bigl(N-k+d\bigr),\qquad k=N,\dots,m.
\end{equation}
Hence, the convolution factor \(e^{-\Lambda_N t}*\dots* e^{-\Lambda_m t}\equiv\mathcal{F}_m(t)\) is trajectory‑independent. For a given final configuration $\vec n=(n_1,\dots,n_d)$ (with $q\!=\!N-m$ total jumps) the population is
\begin{align}
p_{\vec n}(t) &=
\frac{N!\,n_1!\cdots n_d!}{m!}\;
n_{\text{paths}}\;\mathcal{F}_m(t),\quad \ n_{\text{paths}} =\frac{q!}{n_1!\cdots n_d!}.
\end{align}
All $\vec n$ with the same \(m\) therefore share the \emph{same} weight,
\begin{equation}
p_{\vec n}(t)=\frac{\mathcal{C}_m}{\displaystyle\binom{q+d-1}{d-1}}
              \,\mathcal{F}_m(t)
            =\frac{p_m(t)}{\displaystyle\binom{N-m+d-1}{d-1}},
\end{equation}
with the familiar
\(\mathcal{C}_m=N!(N-m)!/m!\) and \(p_m(t)=\mathcal{C}_m\,\mathcal{F}_m(t)\).

For channel~1 we have
\(\hat S_1^\dagger\hat S_1\ket{\vec n,m}=m(n_1+1)\ket{\vec n,m}\).
Using the uniform distribution just derived,
\begin{align}
\mathcal{I}_1(t)
 &=\frac{\Gamma}{d}\sum_{m=0}^{N}\sum_{\vec n}^{q=N-m}
    m(n_1+1)\,p_{\vec n}(t) = \frac{\Gamma}{d}\sum_{m=0}^{N}  m\,p_m(t)\,
   \Bigl[1+\frac{N-m}{d}\Bigr] =\frac{\Gamma}{d}\sum_{m=0}^{N}
   \frac{m\,(N-m+d)}{d}\;p_m(t). 
   \label{eq:I1_d_channels}
\end{align}
By symmetry \(\mathcal{I}_\alpha(t)=\mathcal{I}_1(t)\) for every
\(\alpha\). The total emitted intensity is therefore
\begin{equation}
\mathcal{I}_{\text{tot}}(t)=
\sum_{\alpha=1}^{d}\mathcal{I}_\alpha(t)=\frac{\Gamma}{d}\sum_{m=0}^{N}  m\,(N-m+d)\;p_m(t).
\end{equation}

\subsection{Balanced decay: Peak intensity time}
In the balanced case $\Gamma_\alpha = \Gamma/d$ with  $(\alpha=1,\dots,d),
$, the construction of the trajectory implies that {every trajectory that reaches a fixed Dicke excitation level $k$}
experiences the same non-Hermitian segment rate
\begin{equation}
\Lambda_k \;=\; \frac{\Gamma}{d}\,k\,(N-k+d),
\qquad k=N,N-1,\dots,m.
\label{eq:balanced-Lambda}
\end{equation}

The convolution/Laplace solutions are a cascade in the excitation number $m$: $N \to N-1 \to \cdots \to 0$, where the dwell time in state $k$ is exponential with rate $\Lambda_k$.
Hence, the mean time to {reach} state $m$ is
\begin{equation}
 \bar{T}_m=\sum_{k=m+1}^{N}\frac{1}{\Lambda_k}.
\label{eq:mean-reaching-time}
\end{equation}

Instantaneous intensity is controlled by the quadratic factor $k(N-k+d)$, which is maximized at $k_\star := \left\lfloor\frac{N+d}{2}\right\rfloor$.

In Ref.~\cite{gross_haroche}, the peak time (for a single channel $d=1$) is set as the mean time to reach the maximally radiating state:
\begin{equation}
t_{\rm peak}\;\approx\;\bar{T}_{k_\star}
=\sum_{k=k_\star+1}^{N}\frac{1}{\Lambda_k}
=\frac{d}{\Gamma}\sum_{k=k_\star+1}^{N}\frac{1}{k(N-k+d)}.
\label{eq:tpeak-sum}
\end{equation}
Now use the partial-fraction identity
\begin{equation}
\frac{1}{k(N-k+d)}=\frac{1}{N+d}\left(\frac{1}{k}+\frac{1}{N-k+d}\right),
\label{eq:partial-frac}
\end{equation}
which turns \eqref{eq:tpeak-sum} into harmonic-number sums. Writing $H_n:=\sum_{j=1}^n \frac{1}{j}$,
\begin{equation}
    t_{\rm peak} \approx \frac{d}{\Gamma(N+d)}
\left[
\sum_{k=k_\star+1}^{N}\frac{1}{k}
\;+\;
\sum_{k=k_\star+1}^{N}\frac{1}{N-k+d}
\right] = \frac{d}{\Gamma(N+d)}
\bigg[
\big(H_N-H_{k_\star}\big)
\;+\;
\big(H_{N-k_\star+d-1}-H_{d-1}\big)
\bigg].
\label{eq:tpeak-harmonic}
\end{equation}

Using $k_\star := \left\lfloor\frac{N+d}{2}\right\rfloor$ and the asymptotic expansion of the harmonic number $H_n=\ln n+\gamma_\text{EM}+O(1/n)$, where $\gamma_\text{EM}\approx 0.577$ is the Euler-Mascheroni constant, we obtain the simplification \begin{equation}
\Gamma\,t_{\rm peak}
\;\approx\;
\frac{ d}{N+d}\ln (N/d),
\label{eq:tpeak-final}
\end{equation}
up to $O(d/N)$ and $O(1)$ corrections inside the logarithm. Note that this is precisely the same time obtained from a mean-field treament of the equation $\dot k=-\Lambda_k$.

\section{Stationary state ground state distribution for two channels}
\label{App:TwoChannelSteadyState}

\begin{figure}
    \centering
    \includegraphics[width=0.85\linewidth]{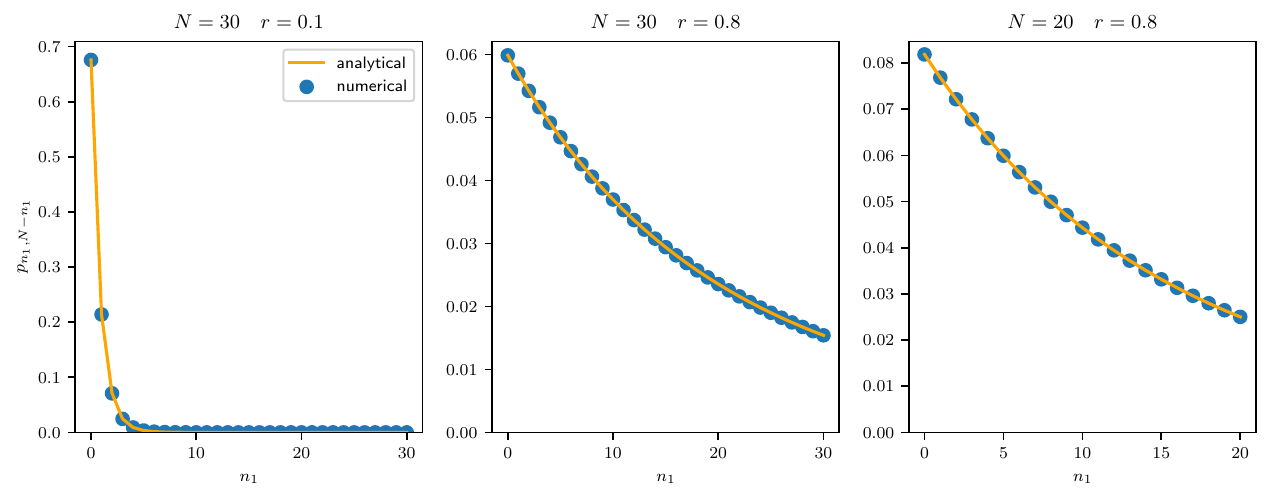}
    \caption{Comparison of numerical results for the \steady-state from rate equation calculation with \steady-state formula in Eq.~\eqref{eq:main_formula}. The two solutions show perfect agreement for representative values of $N$ and $r$.}
    \label{fig:steady_state_check}
\end{figure}

For the unique left–branch path, the Laplace transform is
\begin{equation}
\;p_{N,0}^{(\mathrm{ss})}
   =\frac{N!\,\Gammafunc(1+r)}
          {\Gammafunc(N+1+r)},
\end{equation}
where $\Gammafunc$ denotes the Gamma function and $r=\frac{\Gamma_2}{\Gamma_1}$. This corresponds to the ground state where everything decayed into the first ground state or equivalently, the jump operator $\hat S_1$ applied $N$ times. The next term is obtained from all paths with $N-1$ jumps with $\hat{S}_1$ and $1$ jump with $\hat{S}_2$
\begin{equation}
    p_{N-1,1}^{(\mathrm{ss})}
   =(N-1)! \ r \
     \frac{\Gammafunc(1+r)}{\Gammafunc(N+2r)}
     \left[
       \frac{\Gammafunc(N+1+2r)}{\Gammafunc(N+1+r)}
      -\frac{\Gammafunc(1+2r)}{\Gammafunc(1+r)}
     \right].
\end{equation}
Below is the closed recursive/nested-sum formula that produces every \steady-state population \( p_{N-x,x}^{(\mathrm{ss})} \) for \( x = 0,1,\dots,N \)
\begin{equation}
\label{eq:main_formula}
p_{N-x,x}^{(\mathrm{ss})}
= {(N - x)!}{x!} \cdot r^x \cdot 
  \frac{\Gammafunc(1 + r)}{\Gammafunc(N - (x - 1) + (x + 1)r)} \cdot 
  \sum_{1 \leq L_1 < \cdots < L_x \leq N}
  \prod_{j = 1}^x 
  \frac{
    \Gammafunc(L_j + (j + 1)r - (j - 1))
  }{
    \Gammafunc(L_j + jr - (j - 2))
  },
\end{equation}
which we compare with the numerical solution in Fig.~\ref{fig:steady_state_check}, showing perfect agreement ($\tilde{\Gamma}(\cdot)$ denotes the Gamma function). In particular one finds that for $r=1$ that the distribution is flat $p_{N-x,x}^{(\mathrm{ss})}=1/(N+1)$.

\subsection{Linear expansion of the two-channel \steady{} state around balanced decay}

We will proceed with a linear expansion $r=1+\delta$ with $|\delta|\ll 1$ to derive a logarithmic divergence around the balanced decay point, pointing towards a dissipative phase transition. Let us define the decomposition
\[
p_x(r):=p^{(\mathrm{ss})}_{N-x,x}(r)=A_x(r)\,S_x(r),
\]
where
\begin{align}
A_x(r)&:=(N-x)!\,x!\; r^x\;\frac{\tilde{\Gamma}(1+r)}{\tilde{\Gamma}\!\big(N-(x-1)+(x+1)r\big)},\\
S_x(r)&:=\sum_{1\le L_1<\cdots<L_x\le N}\;
\prod_{j=1}^x
\frac{\tilde{\Gamma}\!\big(L_j+(j+1)r-(j-1)\big)}{\tilde{\Gamma}\!\big(L_j+jr-(j-2)\big)}. \label{a20}
\end{align}
Then
\[
\frac{p_x'(r)}{p_x(r)}=\partial_r\ln A_x(r)+\partial_r\ln S_x(r).
\]

Using $\frac{d}{dr}\ln\Gammafunc(f(r))=\psi(f(r))\,f'(r)$ with the digamma function $\psi$, we get
\begin{equation}
\left.\partial_r\ln A_x(r)\right|_{r=1}
=
x+\psi(2)-(x+1)\psi(N+2).
\label{eq:dlnA}
\end{equation}

For a fixed subset $(L_1,\dots,L_x)$, let us define the product terms in Eq.~(\ref{a20}) as
\[
T(L_1,\dots,L_x;r):=\prod_{j=1}^x
\frac{\tilde{\Gamma}\!\big(L_j+(j+1)r-(j-1)\big)}{\tilde{\Gamma}\!\big(L_j+jr-(j-2)\big)}.
\]
Note, that at $r=1$, $T(\cdot;1)=1$. Moreover,
\begin{align}
\left.\partial_r\ln T\right|_{r=1}
=\sum_{j=1}^x
\left[
(j+1) \ \psi \big(L_j+(j+1)r-(j-1)\big)
-
j\ \psi \big(L_j+jr-(j-2)\big)
\right]_{r=1} =\sum_{j=1}^x \psi(L_j+2),
\end{align}
since the two digamma function arguments coincide at $r=1$ and $(j+1)-j=1$.
Thus,
\[
S_x'(1)=\sum_{1\le L_1<\cdots<L_x\le N}\sum_{j=1}^x \psi(L_j+2).
\]
Each $\ell\in\{1,\dots,N\}$ appears in exactly $\binom{N-1}{x-1}$ subsets of size $x$, so
\[
S_x'(1)=\binom{N-1}{x-1}\sum_{\ell=1}^N \psi(\ell+2),
\qquad
S_x(1)=\binom{N}{x},
\]
hence
\begin{equation}
\left.\partial_r\ln S_x(r)\right|_{r=1}
=
\frac{S_x'(1)}{S_x(1)}
=
\frac{x}{N}\sum_{\ell=1}^N \psi(\ell+2).
\label{eq:dlnS}
\end{equation}

Combining Eqs.~\eqref{eq:dlnA} and \eqref{eq:dlnS}, we first obtain
\begin{align}
\left.\frac{p_x'(r)}{p_x(r)}\right|_{r=1}
&=
x+\psi(2)-(x+1)\psi(N+2)
+\frac{x}{N}\sum_{\ell=1}^N\psi(\ell+2).
\label{eq:pxprime_intermediate}
\end{align}
Using $\psi(n)=H_{n-1}-\gamma_{\mathrm{EM}}$, the Euler--Mascheroni constants cancel explicitly. This gives
\begin{align}
\left.\frac{p_x'(r)}{p_x(r)}\right|_{r=1}
&=
x+H_1-(x+1)H_{N+1}
+\frac{x}{N}\sum_{\ell=1}^N H_{\ell+1}.
\end{align}
With $H_1=1$ and
\[
\sum_{\ell=1}^N H_{\ell+1}=(N+2)\left(H_{N+1}-1\right),
\]
this reduces to
\begin{equation}
\left.\frac{p_x'(r)}{p_x(r)}\right|_{r=1}
=
\left(\frac{2x}{N}-1\right)\left(H_{N+1}-1\right).
\label{eq:pxprime_over_px}
\end{equation}

Using the result for balanced decay, $p_x(1)=1/(N+1)$, the linear expansion $r=1+\delta$ reads
\begin{equation}
p^{(\mathrm{ss})}_{N-x,x}(1+\delta)
=
\frac{1}{N+1}
\left[
1+\delta\left(\frac{2x}{N}-1\right)\big(H_{N+1}-1\big)
\right]
+O(\delta^2).
\label{eq:p_linear_delta}
\end{equation}

\subsection{Order parameter $\bar n_2$ around the balanced decay point}

Recall the definition of the order parameter
\[
\bar n_2(N,r):=\frac{\langle n_2\rangle}{N}
=\frac{1}{N}\sum_{x=0}^N x\,p^{(\mathrm{ss})}_{N-x,x}(r).
\]
Insert \eqref{eq:p_linear_delta}. Using the uniform sums
\[
\sum_{x=0}^N x=\frac{N(N+1)}{2},
\qquad
\sum_{x=0}^N x\left(\frac{2x}{N}-1\right)=\frac{(N+1)(N+2)}{6},
\]
we obtain
\begin{equation}
\bar n_2(N,1+\delta)
=
\frac{1}{2}
+\delta\,\frac{N+2}{6N}\,\big(H_{N+1}-1\big)
+O(\delta^2).
\label{eq:nbar_linear_delta}
\end{equation}
From the standard asymptotic expansion of the harmonic numbers,
\begin{equation}
H_{N+1}=\ln(N)+\gamma_\text{EM}+\mathcal{O}(1/N)
\label{eq:HN1_asymp}
\end{equation}
where $\gamma_\text{EM}$ is the Euler-Mascheroni constant. Using this expansion, the slope grows $\propto \ln N$:
\begin{equation}
\partial_r \bar{n}_2 |_{r=1} = \frac{N+2}{6N}(H_{N+1}-1) \approx \frac{\ln (N)}{6} + O(1).
\end{equation}

\subsection{Analogy to dissipative quantum many-body phase transition for $N\rightarrow \infty$}\label{app:large_N}
\label{App:PhaseTransitionTwoChannel}
\newcommand{\ot}{\tau}

 The \steady{} state distribution becomes analogous to a many-body phase transition in the large-\(N\) limit in terms of the collective decay rate ratio $r$: the susceptibility defined as \(\chi=\partial \bar{n}_2/\partial r\sim \ln N\) diverges at \(r=1\), indicating a first-order transition that sharpens logarithmically with $N$. Intuitively, because \(\ot\sim (\ln N)/\Gamma_{\max}\), the faster channel “wins” (referred to as "winner-takes-all" in Ref.~\cite{mok2025}) unless the rates are degenerate, capturing a genuine dissipative many-body phase transition arising purely from competing decay channels under the constraint \(\sum_\alpha \langle n_\alpha \rangle=N\) (i.e. system in the ground state).

\section{Recursive equations for multichannel Dicke superradiance}

The recursive rate equations below provide an efficient numerical cross-check of the symbolic trajectory solution. They exploit the same permutation symmetry but should be distinguished from the trajectory construction: the former propagate the symmetry-reduced populations numerically, whereas the latter evaluates the ordered decay paths analytically and yields the finite convolution/Laplace expressions used in the main text.

In Dicke superradiance, starting from the fully excited state, the density matrix remains diagonal in the Dicke basis at all times. Thus, the system can be solved based on a system of coupled recursive rate equations. The recursive equation for multichannel Dicke superradiance reads
\begin{align} \label{eq:rate}
\frac{d}{dt}\,p_{n_1,\cdots,n_d}(t)
&=
-\Lambda_m\,
  p_{n_1,\cdots,n_d}(t) + \sum_{\alpha=1}^d\;\Gamma_\alpha\,(m+1)\,n_\alpha\;p_{\,n_1,\cdots,n_\alpha-1,\cdots,n_d}(t),
\end{align}
where $m = N - \sum_{\alpha=1}^d n_\alpha$, $\Lambda_m = \sum_{\alpha=1}^d m \Gamma_\alpha(n_\alpha+1)$ and $p_{n_1,\cdots,n_d}(0)=\delta_{n_1,0} \cdots \delta_{n_d,0}$. In particular for $d=2$, the recursive equation for two-channel Dicke superradiance reads
\begin{align}
\frac{d}{dt}\,p_{n_1,n_2}(t)
&=
-\Lambda_m\,
  p_{n_1,n_2}(t)
+ \Gamma_1\,(m+1)\,n_1\,p_{n_1-1,n_2}(t)
+ \Gamma_2\,(m+1)\,n_2\,p_{n_1,n_2-1}(t),
\end{align}
where $m = N - n_1 - n_2$, $\Lambda_m = m\left[\,\Gamma_1(n_1+1) + \Gamma_2(n_2+1)\,\right]$ and with initial condition $p_{n_1,n_2}(0) = \delta_{n_1,0}\,\delta_{n_2,0}$.


\begin{thebibliography}{43}%
\makeatletter
\providecommand \@ifxundefined [1]{%
 \@ifx{#1\undefined}
}%
\providecommand \@ifnum [1]{%
 \ifnum #1\expandafter \@firstoftwo
 \else \expandafter \@secondoftwo
 \fi
}%
\providecommand \@ifx [1]{%
 \ifx #1\expandafter \@firstoftwo
 \else \expandafter \@secondoftwo
 \fi
}%
\providecommand \natexlab [1]{#1}%
\providecommand \enquote  [1]{``#1''}%
\providecommand \bibnamefont  [1]{#1}%
\providecommand \bibfnamefont [1]{#1}%
\providecommand \citenamefont [1]{#1}%
\providecommand \href@noop [0]{\@secondoftwo}%
\providecommand \href [0]{\begingroup \@sanitize@url \@href}%
\providecommand \@href[1]{\@@startlink{#1}\@@href}%
\providecommand \@@href[1]{\endgroup#1\@@endlink}%
\providecommand \@sanitize@url [0]{\catcode `\\12\catcode `\$12\catcode `\&12\catcode `\#12\catcode `\^12\catcode `\_12\catcode `\%12\relax}%
\providecommand \@@startlink[1]{}%
\providecommand \@@endlink[0]{}%
\providecommand \url  [0]{\begingroup\@sanitize@url \@url }%
\providecommand \@url [1]{\endgroup\@href {#1}{\urlprefix }}%
\providecommand \urlprefix  [0]{URL }%
\providecommand \Eprint [0]{\href }%
\providecommand \doibase [0]{https://doi.org/}%
\providecommand \selectlanguage [0]{\@gobble}%
\providecommand \bibinfo  [0]{\@secondoftwo}%
\providecommand \bibfield  [0]{\@secondoftwo}%
\providecommand \translation [1]{[#1]}%
\providecommand \BibitemOpen [0]{}%
\providecommand \bibitemStop [0]{}%
\providecommand \bibitemNoStop [0]{.\EOS\space}%
\providecommand \EOS [0]{\spacefactor3000\relax}%
\providecommand \BibitemShut  [1]{\csname bibitem#1\endcsname}%
\let\auto@bib@innerbib\@empty
\bibitem [{\citenamefont {Dicke}(1954)}]{dicke1954}%
  \BibitemOpen
  \bibfield  {author} {\bibinfo {author} {\bibfnamefont {R.~H.}\ \bibnamefont {Dicke}},\ }\bibfield  {title} {\bibinfo {title} {Coherence in spontaneous radiation processes},\ }\href {https://doi.org/10.1103/PhysRev.93.99} {\bibfield  {journal} {\bibinfo  {journal} {Phys. Rev.}\ }\textbf {\bibinfo {volume} {93}},\ \bibinfo {pages} {99} (\bibinfo {year} {1954})}\BibitemShut {NoStop}%
\bibitem [{\citenamefont {Temnov}\ and\ \citenamefont {Woggon}(2005)}]{temnov2005}%
  \BibitemOpen
  \bibfield  {author} {\bibinfo {author} {\bibfnamefont {V.~V.}\ \bibnamefont {Temnov}}\ and\ \bibinfo {author} {\bibfnamefont {U.}~\bibnamefont {Woggon}},\ }\bibfield  {title} {\bibinfo {title} {Superradiance and subradiance in an inhomogeneously broadened ensemble of two-level systems coupled to a low-{Q} cavity},\ }\href {https://doi.org/10.1103/PhysRevLett.95.243602} {\bibfield  {journal} {\bibinfo  {journal} {Phys. Rev. Lett.}\ }\textbf {\bibinfo {volume} {95}},\ \bibinfo {pages} {243602} (\bibinfo {year} {2005})}\BibitemShut {NoStop}%
\bibitem [{\citenamefont {Scully}\ and\ \citenamefont {Svidzinsky}(2009)}]{Scully2009super}%
  \BibitemOpen
  \bibfield  {author} {\bibinfo {author} {\bibfnamefont {M.~O.}\ \bibnamefont {Scully}}\ and\ \bibinfo {author} {\bibfnamefont {A.~A.}\ \bibnamefont {Svidzinsky}},\ }\bibfield  {title} {\bibinfo {title} {The super of superradiance},\ }\href {https://doi.org/10.1126/science.1176695} {\bibfield  {journal} {\bibinfo  {journal} {Science}\ }\textbf {\bibinfo {volume} {325}},\ \bibinfo {pages} {1510} (\bibinfo {year} {2009})}\BibitemShut {NoStop}%
\bibitem [{\citenamefont {Lee}(1977{\natexlab{a}})}]{tsung1977Part1}%
  \BibitemOpen
  \bibfield  {author} {\bibinfo {author} {\bibfnamefont {C.~T.}\ \bibnamefont {Lee}},\ }\bibfield  {title} {\bibinfo {title} {Exact solution of the superradiance master equation. {}. {C}omplete initial excitation},\ }\href {https://doi.org/10.1103/PhysRevA.15.2019} {\bibfield  {journal} {\bibinfo  {journal} {Phys. Rev. A}\ }\textbf {\bibinfo {volume} {15}},\ \bibinfo {pages} {2019} (\bibinfo {year} {1977}{\natexlab{a}})}\BibitemShut {NoStop}%
\bibitem [{\citenamefont {Lee}(1977{\natexlab{b}})}]{tsung1977Part2}%
  \BibitemOpen
  \bibfield  {author} {\bibinfo {author} {\bibfnamefont {C.~T.}\ \bibnamefont {Lee}},\ }\bibfield  {title} {\bibinfo {title} {Exact solution of the superradiance master equation. {II}. {A}rbitrary initial excitation},\ }\href {https://doi.org/10.1103/PhysRevA.16.301} {\bibfield  {journal} {\bibinfo  {journal} {Phys. Rev. A}\ }\textbf {\bibinfo {volume} {16}},\ \bibinfo {pages} {301} (\bibinfo {year} {1977}{\natexlab{b}})}\BibitemShut {NoStop}%
\bibitem [{\citenamefont {Holzinger}\ and\ \citenamefont {Genes}(2025)}]{holzinger2025a}%
  \BibitemOpen
  \bibfield  {author} {\bibinfo {author} {\bibfnamefont {R.}~\bibnamefont {Holzinger}}\ and\ \bibinfo {author} {\bibfnamefont {C.}~\bibnamefont {Genes}},\ }\bibfield  {title} {\bibinfo {title} {A compact analytical solution of the {D}icke superradiance master equation via residue calculus},\ }\href {https://doi.org/doi:10.1515/zna-2025-0179} {\bibfield  {journal} {\bibinfo  {journal} {Zeitschrift für Naturforschung A}\ }\textbf {\bibinfo {volume} {80}},\ \bibinfo {pages} {673} (\bibinfo {year} {2025})}\BibitemShut {NoStop}%
\bibitem [{\citenamefont {Holzinger}\ \emph {et~al.}(2026)\citenamefont {Holzinger}, \citenamefont {Bassler}, \citenamefont {Lyne}, \citenamefont {Jimenez}, \citenamefont {Gohsrich},\ and\ \citenamefont {Genes}}]{holzinger2025b}%
  \BibitemOpen
  \bibfield  {author} {\bibinfo {author} {\bibfnamefont {R.}~\bibnamefont {Holzinger}}, \bibinfo {author} {\bibfnamefont {N.~S.}\ \bibnamefont {Bassler}}, \bibinfo {author} {\bibfnamefont {J.}~\bibnamefont {Lyne}}, \bibinfo {author} {\bibfnamefont {F.~G.}\ \bibnamefont {Jimenez}}, \bibinfo {author} {\bibfnamefont {J.~T.}\ \bibnamefont {Gohsrich}},\ and\ \bibinfo {author} {\bibfnamefont {C.}~\bibnamefont {Genes}},\ }\bibfield  {title} {\bibinfo {title} {Solving {D}icke superradiance analytically: A compendium of methods},\ }\href {https://doi.org/10.1103/9p44-kwfl} {\bibfield  {journal} {\bibinfo  {journal} {Phys. Rev. A}\ }\textbf {\bibinfo {volume} {113}},\ \bibinfo {pages} {043718} (\bibinfo {year} {2026})}\BibitemShut {NoStop}%
\bibitem [{\citenamefont {Sutherland}\ and\ \citenamefont {Robicheaux}(2017)}]{sutherland2017superradiance}%
  \BibitemOpen
  \bibfield  {author} {\bibinfo {author} {\bibfnamefont {R.~T.}\ \bibnamefont {Sutherland}}\ and\ \bibinfo {author} {\bibfnamefont {F.}~\bibnamefont {Robicheaux}},\ }\bibfield  {title} {\bibinfo {title} {Superradiance in inverted multilevel atomic clouds},\ }\href {https://doi.org/10.1103/PhysRevA.95.033839} {\bibfield  {journal} {\bibinfo  {journal} {Phys. Rev. A}\ }\textbf {\bibinfo {volume} {95}},\ \bibinfo {pages} {033839} (\bibinfo {year} {2017})}\BibitemShut {NoStop}%
\bibitem [{\citenamefont {Pi\~neiro Orioli}\ \emph {et~al.}(2022)\citenamefont {Pi\~neiro Orioli}, \citenamefont {Thompson},\ and\ \citenamefont {Rey}}]{Orioli2022emergent}%
  \BibitemOpen
  \bibfield  {author} {\bibinfo {author} {\bibfnamefont {A.}~\bibnamefont {Pi\~neiro Orioli}}, \bibinfo {author} {\bibfnamefont {J.~K.}\ \bibnamefont {Thompson}},\ and\ \bibinfo {author} {\bibfnamefont {A.~M.}\ \bibnamefont {Rey}},\ }\bibfield  {title} {\bibinfo {title} {Emergent dark states from superradiant dynamics in multilevel atoms in a cavity},\ }\href {https://doi.org/10.1103/PhysRevX.12.011054} {\bibfield  {journal} {\bibinfo  {journal} {Phys. Rev. X}\ }\textbf {\bibinfo {volume} {12}},\ \bibinfo {pages} {011054} (\bibinfo {year} {2022})}\BibitemShut {NoStop}%
\bibitem [{\citenamefont {Masson}\ \emph {et~al.}(2024)\citenamefont {Masson}, \citenamefont {Covey}, \citenamefont {Will},\ and\ \citenamefont {Asenjo-Garcia}}]{masson2024dicke}%
  \BibitemOpen
  \bibfield  {author} {\bibinfo {author} {\bibfnamefont {S.~J.}\ \bibnamefont {Masson}}, \bibinfo {author} {\bibfnamefont {J.~P.}\ \bibnamefont {Covey}}, \bibinfo {author} {\bibfnamefont {S.}~\bibnamefont {Will}},\ and\ \bibinfo {author} {\bibfnamefont {A.}~\bibnamefont {Asenjo-Garcia}},\ }\bibfield  {title} {\bibinfo {title} {{D}icke superradiance in ordered arrays of multilevel atoms},\ }\href {https://doi.org/10.1103/PRXQuantum.5.010344} {\bibfield  {journal} {\bibinfo  {journal} {PRX Quantum}\ }\textbf {\bibinfo {volume} {5}},\ \bibinfo {pages} {010344} (\bibinfo {year} {2024})}\BibitemShut {NoStop}%
\bibitem [{\citenamefont {Mok}\ \emph {et~al.}(2025)\citenamefont {Mok}, \citenamefont {Masson}, \citenamefont {Stamper-Kurn}, \citenamefont {Zelevinsky},\ and\ \citenamefont {Asenjo-Garcia}}]{mok2025}%
  \BibitemOpen
  \bibfield  {author} {\bibinfo {author} {\bibfnamefont {W.-K.}\ \bibnamefont {Mok}}, \bibinfo {author} {\bibfnamefont {S.~J.}\ \bibnamefont {Masson}}, \bibinfo {author} {\bibfnamefont {D.~M.}\ \bibnamefont {Stamper-Kurn}}, \bibinfo {author} {\bibfnamefont {T.}~\bibnamefont {Zelevinsky}},\ and\ \bibinfo {author} {\bibfnamefont {A.}~\bibnamefont {Asenjo-Garcia}},\ }\bibfield  {title} {\bibinfo {title} {Ground-state selection via many-body superradiant decay},\ }\href {https://doi.org/10.1103/PhysRevResearch.7.L022015} {\bibfield  {journal} {\bibinfo  {journal} {Phys. Rev. Res.}\ }\textbf {\bibinfo {volume} {7}},\ \bibinfo {pages} {L022015} (\bibinfo {year} {2025})}\BibitemShut {NoStop}%
\bibitem [{\citenamefont {Sivan}\ and\ \citenamefont {Orenstein}(2025)}]{sivan2025multilevel}%
  \BibitemOpen
  \bibfield  {author} {\bibinfo {author} {\bibfnamefont {A.}~\bibnamefont {Sivan}}\ and\ \bibinfo {author} {\bibfnamefont {M.}~\bibnamefont {Orenstein}},\ }\bibfield  {title} {\bibinfo {title} {Adding photonic entanglement to superradiance by using multilevel atoms},\ }\href {https://doi.org/10.1103/yv7y-pyjk} {\bibfield  {journal} {\bibinfo  {journal} {Phys. Rev. Res.}\ }\textbf {\bibinfo {volume} {7}},\ \bibinfo {pages} {033170} (\bibinfo {year} {2025})}\BibitemShut {NoStop}%
\bibitem [{\citenamefont {Chase}\ and\ \citenamefont {Geremia}(2008)}]{Chase2008}%
  \BibitemOpen
  \bibfield  {author} {\bibinfo {author} {\bibfnamefont {B.~A.}\ \bibnamefont {Chase}}\ and\ \bibinfo {author} {\bibfnamefont {J.~M.}\ \bibnamefont {Geremia}},\ }\bibfield  {title} {\bibinfo {title} {Collective processes of an ensemble of spin-1/2 particles},\ }\href {https://doi.org/10.1103/PhysRevA.78.052101} {\bibfield  {journal} {\bibinfo  {journal} {Physical Review A}\ }\textbf {\bibinfo {volume} {78}},\ \bibinfo {pages} {052101} (\bibinfo {year} {2008})}\BibitemShut {NoStop}%
\bibitem [{\citenamefont {Baragiola}\ \emph {et~al.}(2010)\citenamefont {Baragiola}, \citenamefont {Chase},\ and\ \citenamefont {Geremia}}]{Baragiola2010}%
  \BibitemOpen
  \bibfield  {author} {\bibinfo {author} {\bibfnamefont {B.~Q.}\ \bibnamefont {Baragiola}}, \bibinfo {author} {\bibfnamefont {B.~A.}\ \bibnamefont {Chase}},\ and\ \bibinfo {author} {\bibfnamefont {J.~M.}\ \bibnamefont {Geremia}},\ }\bibfield  {title} {\bibinfo {title} {Collective uncertainty in partially polarized and partially decohered spin-1/2 systems},\ }\href {https://doi.org/10.1103/PhysRevA.81.032104} {\bibfield  {journal} {\bibinfo  {journal} {Physical Review A}\ }\textbf {\bibinfo {volume} {81}},\ \bibinfo {pages} {032104} (\bibinfo {year} {2010})}\BibitemShut {NoStop}%
\bibitem [{\citenamefont {Xu}\ \emph {et~al.}(2013)\citenamefont {Xu}, \citenamefont {Tieri},\ and\ \citenamefont {Holland}}]{Xu2013}%
  \BibitemOpen
  \bibfield  {author} {\bibinfo {author} {\bibfnamefont {M.}~\bibnamefont {Xu}}, \bibinfo {author} {\bibfnamefont {D.~A.}\ \bibnamefont {Tieri}},\ and\ \bibinfo {author} {\bibfnamefont {M.~J.}\ \bibnamefont {Holland}},\ }\bibfield  {title} {\bibinfo {title} {Simulating open quantum systems by applying {SU}(4) to quantum master equations},\ }\href {https://doi.org/10.1103/PhysRevA.87.062101} {\bibfield  {journal} {\bibinfo  {journal} {Physical Review A}\ }\textbf {\bibinfo {volume} {87}},\ \bibinfo {pages} {062101} (\bibinfo {year} {2013})}\BibitemShut {NoStop}%
\bibitem [{\citenamefont {Zhang}\ \emph {et~al.}(2018)\citenamefont {Zhang}, \citenamefont {Zhang},\ and\ \citenamefont {M{\o}lmer}}]{Zhang2018}%
  \BibitemOpen
  \bibfield  {author} {\bibinfo {author} {\bibfnamefont {Y.}~\bibnamefont {Zhang}}, \bibinfo {author} {\bibfnamefont {Y.-X.}\ \bibnamefont {Zhang}},\ and\ \bibinfo {author} {\bibfnamefont {K.}~\bibnamefont {M{\o}lmer}},\ }\bibfield  {title} {\bibinfo {title} {Monte-carlo simulations of superradiant lasing},\ }\href {https://doi.org/10.1088/1367-2630/aaec36} {\bibfield  {journal} {\bibinfo  {journal} {New Journal of Physics}\ }\textbf {\bibinfo {volume} {20}},\ \bibinfo {pages} {112001} (\bibinfo {year} {2018})}\BibitemShut {NoStop}%
\bibitem [{\citenamefont {Lloyd}\ \emph {et~al.}(2026)\citenamefont {Lloyd}, \citenamefont {Ziolkowska},\ and\ \citenamefont {Keeling}}]{Lloyd2026}%
  \BibitemOpen
  \bibfield  {author} {\bibinfo {author} {\bibfnamefont {E.~W.}\ \bibnamefont {Lloyd}}, \bibinfo {author} {\bibfnamefont {A.~A.}\ \bibnamefont {Ziolkowska}},\ and\ \bibinfo {author} {\bibfnamefont {J.}~\bibnamefont {Keeling}},\ }\href {https://arxiv.org/abs/2605.11103} {\bibinfo {title} {Permutation-symmetric quantum trajectories}} (\bibinfo {year} {2026}),\ \Eprint {https://arxiv.org/abs/2605.11103} {arXiv:2605.11103 [quant-ph]} \BibitemShut {NoStop}%
\bibitem [{\citenamefont {Mølmer}\ \emph {et~al.}(1993)\citenamefont {Mølmer}, \citenamefont {Castin},\ and\ \citenamefont {Dalibard}}]{dalibard1992}%
  \BibitemOpen
  \bibfield  {author} {\bibinfo {author} {\bibfnamefont {K.}~\bibnamefont {Mølmer}}, \bibinfo {author} {\bibfnamefont {Y.}~\bibnamefont {Castin}},\ and\ \bibinfo {author} {\bibfnamefont {J.}~\bibnamefont {Dalibard}},\ }\bibfield  {title} {\bibinfo {title} {Monte {Carlo} wave-function method in quantum optics},\ }\href {https://doi.org/10.1364/JOSAB.10.000524} {\bibfield  {journal} {\bibinfo  {journal} {Journal of the Optical Society of America B}\ }\textbf {\bibinfo {volume} {10}},\ \bibinfo {pages} {524} (\bibinfo {year} {1993})}\BibitemShut {NoStop}%
\bibitem [{\citenamefont {Carmichael}(1993)}]{Carmichael1993}%
  \BibitemOpen
  \bibfield  {author} {\bibinfo {author} {\bibfnamefont {H.~J.}\ \bibnamefont {Carmichael}},\ }\href {https://doi.org/10.1007/978-3-540-47620-7} {\emph {\bibinfo {title} {An Open Systems Approach to Quantum Optics}}},\ \bibinfo {series} {Lecture Notes in Physics}, Vol.\ \bibinfo {volume} {m18}\ (\bibinfo  {publisher} {Springer},\ \bibinfo {address} {Berlin},\ \bibinfo {year} {1993})\BibitemShut {NoStop}%
\bibitem [{\citenamefont {Plenio}\ and\ \citenamefont {Knight}(1998)}]{Plenio1998}%
  \BibitemOpen
  \bibfield  {author} {\bibinfo {author} {\bibfnamefont {M.~B.}\ \bibnamefont {Plenio}}\ and\ \bibinfo {author} {\bibfnamefont {P.~L.}\ \bibnamefont {Knight}},\ }\bibfield  {title} {\bibinfo {title} {The quantum-jump approach to dissipative dynamics in quantum optics},\ }\href {https://doi.org/10.1103/RevModPhys.70.101} {\bibfield  {journal} {\bibinfo  {journal} {Rev. Mod. Phys.}\ }\textbf {\bibinfo {volume} {70}},\ \bibinfo {pages} {101} (\bibinfo {year} {1998})}\BibitemShut {NoStop}%
\bibitem [{\citenamefont {Park}\ \emph {et~al.}(2022)\citenamefont {Park}, \citenamefont {Trautmann}, \citenamefont {\ifmmode \check{S}\else \v{S}\fi{}anti\ifmmode~\acute{c}\else \'{c}\fi{}}, \citenamefont {Kl\"usener}, \citenamefont {Heinz}, \citenamefont {Bloch},\ and\ \citenamefont {Blatt}}]{park2022cavity}%
  \BibitemOpen
  \bibfield  {author} {\bibinfo {author} {\bibfnamefont {A.~J.}\ \bibnamefont {Park}}, \bibinfo {author} {\bibfnamefont {J.}~\bibnamefont {Trautmann}}, \bibinfo {author} {\bibfnamefont {N.}~\bibnamefont {\ifmmode \check{S}\else \v{S}\fi{}anti\ifmmode~\acute{c}\else \'{c}\fi{}}}, \bibinfo {author} {\bibfnamefont {V.}~\bibnamefont {Kl\"usener}}, \bibinfo {author} {\bibfnamefont {A.}~\bibnamefont {Heinz}}, \bibinfo {author} {\bibfnamefont {I.}~\bibnamefont {Bloch}},\ and\ \bibinfo {author} {\bibfnamefont {S.}~\bibnamefont {Blatt}},\ }\bibfield  {title} {\bibinfo {title} {Cavity-enhanced optical lattices for scaling neutral atom quantum technologies to higher qubit numbers},\ }\href {https://doi.org/10.1103/PRXQuantum.3.030314} {\bibfield  {journal} {\bibinfo  {journal} {PRX Quantum}\ }\textbf {\bibinfo {volume} {3}},\ \bibinfo {pages} {030314} (\bibinfo {year} {2022})}\BibitemShut {NoStop}%
\bibitem [{SM()}]{SM}%
  \BibitemOpen
  \href@noop {} {}\bibinfo {note} {See Supplemental Material at [URL will be inserted by publisher] for details on the adiabatic elimination in the bad-cavity limit, solving Dicke superradiance for more than two channels, balanced decay, steady-state ground state distribution for two channels, analogy to dissipative quantum many body phase transition as well as mean field and stochastic simulations for multichannel Dicke superradiance.}\BibitemShut {Stop}%
\bibitem [{\citenamefont {Bohnet}\ \emph {et~al.}(2012)\citenamefont {Bohnet}, \citenamefont {Chen}, \citenamefont {Weiner}, \citenamefont {Meiser}, \citenamefont {Holland},\ and\ \citenamefont {Thompson}}]{Bohnet2012}%
  \BibitemOpen
  \bibfield  {author} {\bibinfo {author} {\bibfnamefont {J.~G.}\ \bibnamefont {Bohnet}}, \bibinfo {author} {\bibfnamefont {Z.}~\bibnamefont {Chen}}, \bibinfo {author} {\bibfnamefont {J.~M.}\ \bibnamefont {Weiner}}, \bibinfo {author} {\bibfnamefont {D.}~\bibnamefont {Meiser}}, \bibinfo {author} {\bibfnamefont {M.~J.}\ \bibnamefont {Holland}},\ and\ \bibinfo {author} {\bibfnamefont {J.~K.}\ \bibnamefont {Thompson}},\ }\bibfield  {title} {\bibinfo {title} {A steady-state superradiant laser with less than one intracavity photon},\ }\href {https://doi.org/10.1038/nature10920} {\bibfield  {journal} {\bibinfo  {journal} {Nature}\ }\textbf {\bibinfo {volume} {484}},\ \bibinfo {pages} {78} (\bibinfo {year} {2012})}\BibitemShut {NoStop}%
\bibitem [{\citenamefont {Norcia}\ and\ \citenamefont {Thompson}(2016)}]{norcia2016prx}%
  \BibitemOpen
  \bibfield  {author} {\bibinfo {author} {\bibfnamefont {M.~A.}\ \bibnamefont {Norcia}}\ and\ \bibinfo {author} {\bibfnamefont {J.~K.}\ \bibnamefont {Thompson}},\ }\bibfield  {title} {\bibinfo {title} {Cold-strontium laser in the superradiant crossover regime},\ }\href {https://doi.org/10.1103/PhysRevX.6.011025} {\bibfield  {journal} {\bibinfo  {journal} {Phys. Rev. X}\ }\textbf {\bibinfo {volume} {6}},\ \bibinfo {pages} {011025} (\bibinfo {year} {2016})}\BibitemShut {NoStop}%
\bibitem [{\citenamefont {Norcia}\ \emph {et~al.}(2016)\citenamefont {Norcia}, \citenamefont {Winchester}, \citenamefont {Cline},\ and\ \citenamefont {Thompson}}]{norcia2016superradiance}%
  \BibitemOpen
  \bibfield  {author} {\bibinfo {author} {\bibfnamefont {M.~A.}\ \bibnamefont {Norcia}}, \bibinfo {author} {\bibfnamefont {M.~N.}\ \bibnamefont {Winchester}}, \bibinfo {author} {\bibfnamefont {J.~R.~K.}\ \bibnamefont {Cline}},\ and\ \bibinfo {author} {\bibfnamefont {J.~K.}\ \bibnamefont {Thompson}},\ }\bibfield  {title} {\bibinfo {title} {{Superradiance on the millihertz linewidth {s}trontium clock transition}},\ }\href {https://advances.sciencemag.org/content/2/10/e1601231} {\bibfield  {journal} {\bibinfo  {journal} {Science Advances}\ }\textbf {\bibinfo {volume} {2}} (\bibinfo {year} {2016})}\BibitemShut {NoStop}%
\bibitem [{\citenamefont {Norcia}\ \emph {et~al.}(2018)\citenamefont {Norcia}, \citenamefont {Cline}, \citenamefont {Muniz}, \citenamefont {Robinson}, \citenamefont {Hutson}, \citenamefont {Goban}, \citenamefont {Marti}, \citenamefont {Ye},\ and\ \citenamefont {Thompson}}]{norcia2018prx}%
  \BibitemOpen
  \bibfield  {author} {\bibinfo {author} {\bibfnamefont {M.~A.}\ \bibnamefont {Norcia}}, \bibinfo {author} {\bibfnamefont {J.~R.~K.}\ \bibnamefont {Cline}}, \bibinfo {author} {\bibfnamefont {J.~A.}\ \bibnamefont {Muniz}}, \bibinfo {author} {\bibfnamefont {J.~M.}\ \bibnamefont {Robinson}}, \bibinfo {author} {\bibfnamefont {R.~B.}\ \bibnamefont {Hutson}}, \bibinfo {author} {\bibfnamefont {A.}~\bibnamefont {Goban}}, \bibinfo {author} {\bibfnamefont {G.~E.}\ \bibnamefont {Marti}}, \bibinfo {author} {\bibfnamefont {J.}~\bibnamefont {Ye}},\ and\ \bibinfo {author} {\bibfnamefont {J.~K.}\ \bibnamefont {Thompson}},\ }\bibfield  {title} {\bibinfo {title} {Frequency measurements of superradiance from the strontium clock transition},\ }\href {https://doi.org/10.1103/PhysRevX.8.021036} {\bibfield  {journal} {\bibinfo  {journal} {Phys. Rev. X}\ }\textbf {\bibinfo {volume} {8}},\ \bibinfo {pages} {021036} (\bibinfo {year} {2018})}\BibitemShut {NoStop}%
\bibitem [{\citenamefont {Mlynek}\ \emph {et~al.}(2014)\citenamefont {Mlynek}, \citenamefont {Abdumalikov~Jr.}, \citenamefont {Eichler},\ and\ \citenamefont {Wallraff}}]{mlynek2014}%
  \BibitemOpen
  \bibfield  {author} {\bibinfo {author} {\bibfnamefont {J.~A.}\ \bibnamefont {Mlynek}}, \bibinfo {author} {\bibfnamefont {A.~A.}\ \bibnamefont {Abdumalikov~Jr.}}, \bibinfo {author} {\bibfnamefont {C.}~\bibnamefont {Eichler}},\ and\ \bibinfo {author} {\bibfnamefont {A.}~\bibnamefont {Wallraff}},\ }\bibfield  {title} {\bibinfo {title} {Observation of {D}icke superradiance for two artificial atoms in a cavity with high decay rate},\ }\href {https://doi.org/10.1038/ncomms6186} {\bibfield  {journal} {\bibinfo  {journal} {Nat. Commun.}\ }\textbf {\bibinfo {volume} {5}},\ \bibinfo {pages} {5186} (\bibinfo {year} {2014})}\BibitemShut {NoStop}%
\bibitem [{\citenamefont {Brekenfeld}\ \emph {et~al.}(2020)\citenamefont {Brekenfeld}, \citenamefont {Niemietz}, \citenamefont {Christesen},\ and\ \citenamefont {Rempe}}]{Brekenfeld2020crossed}%
  \BibitemOpen
  \bibfield  {author} {\bibinfo {author} {\bibfnamefont {M.}~\bibnamefont {Brekenfeld}}, \bibinfo {author} {\bibfnamefont {D.}~\bibnamefont {Niemietz}}, \bibinfo {author} {\bibfnamefont {J.~D.}\ \bibnamefont {Christesen}},\ and\ \bibinfo {author} {\bibfnamefont {G.}~\bibnamefont {Rempe}},\ }\bibfield  {title} {\bibinfo {title} {A quantum network node with crossed optical fibre cavities},\ }\href {https://doi.org/10.1038/s41567-020-0855-3} {\bibfield  {journal} {\bibinfo  {journal} {Nature Physics}\ }\textbf {\bibinfo {volume} {16}},\ \bibinfo {pages} {647} (\bibinfo {year} {2020})}\BibitemShut {NoStop}%
\bibitem [{\citenamefont {Goban}\ \emph {et~al.}(2015)\citenamefont {Goban}, \citenamefont {Hung}, \citenamefont {Hood}, \citenamefont {Yu}, \citenamefont {Muniz}, \citenamefont {Lee}, \citenamefont {Martin}, \citenamefont {McClung}, \citenamefont {Choi}, \citenamefont {Chan}, \citenamefont {Painter},\ and\ \citenamefont {Kimble}}]{Goban2015PRL}%
  \BibitemOpen
  \bibfield  {author} {\bibinfo {author} {\bibfnamefont {A.}~\bibnamefont {Goban}}, \bibinfo {author} {\bibfnamefont {C.-L.}\ \bibnamefont {Hung}}, \bibinfo {author} {\bibfnamefont {J.~D.}\ \bibnamefont {Hood}}, \bibinfo {author} {\bibfnamefont {S.-P.}\ \bibnamefont {Yu}}, \bibinfo {author} {\bibfnamefont {J.~A.}\ \bibnamefont {Muniz}}, \bibinfo {author} {\bibfnamefont {J.~H.}\ \bibnamefont {Lee}}, \bibinfo {author} {\bibfnamefont {M.~J.}\ \bibnamefont {Martin}}, \bibinfo {author} {\bibfnamefont {A.}~\bibnamefont {McClung}}, \bibinfo {author} {\bibfnamefont {K.~S.}\ \bibnamefont {Choi}}, \bibinfo {author} {\bibfnamefont {J.}~\bibnamefont {Chan}}, \bibinfo {author} {\bibfnamefont {O.}~\bibnamefont {Painter}},\ and\ \bibinfo {author} {\bibfnamefont {H.~J.}\ \bibnamefont {Kimble}},\ }\bibfield  {title} {\bibinfo {title} {Superradiance for atoms trapped along a photonic crystal waveguide},\ }\href {https://doi.org/10.1103/PhysRevLett.115.063601} {\bibfield  {journal} {\bibinfo  {journal} {Phys. Rev. Lett.}\
  }\textbf {\bibinfo {volume} {115}},\ \bibinfo {pages} {063601} (\bibinfo {year} {2015})}\BibitemShut {NoStop}%
\bibitem [{\citenamefont {Solano}\ \emph {et~al.}(2017)\citenamefont {Solano}, \citenamefont {Barberis-Blostein}, \citenamefont {Fatemi}, \citenamefont {Orozco},\ and\ \citenamefont {Rolston}}]{Solano2017superradiance}%
  \BibitemOpen
  \bibfield  {author} {\bibinfo {author} {\bibfnamefont {P.}~\bibnamefont {Solano}}, \bibinfo {author} {\bibfnamefont {P.}~\bibnamefont {Barberis-Blostein}}, \bibinfo {author} {\bibfnamefont {F.~K.}\ \bibnamefont {Fatemi}}, \bibinfo {author} {\bibfnamefont {L.~A.}\ \bibnamefont {Orozco}},\ and\ \bibinfo {author} {\bibfnamefont {S.~L.}\ \bibnamefont {Rolston}},\ }\bibfield  {title} {\bibinfo {title} {Super-radiance reveals infinite-range dipole interactions through a nanofiber},\ }\href {https://doi.org/10.1038/s41467-017-01994-3} {\bibfield  {journal} {\bibinfo  {journal} {Nature Communications}\ }\textbf {\bibinfo {volume} {8}},\ \bibinfo {pages} {1857} (\bibinfo {year} {2017})}\BibitemShut {NoStop}%
\bibitem [{\citenamefont {Sheremet}\ \emph {et~al.}(2023)\citenamefont {Sheremet}, \citenamefont {Petrov}, \citenamefont {Iorsh}, \citenamefont {Poshakinskiy},\ and\ \citenamefont {Poddubny}}]{sheremet2023waveguide}%
  \BibitemOpen
  \bibfield  {author} {\bibinfo {author} {\bibfnamefont {A.~S.}\ \bibnamefont {Sheremet}}, \bibinfo {author} {\bibfnamefont {M.~I.}\ \bibnamefont {Petrov}}, \bibinfo {author} {\bibfnamefont {I.~V.}\ \bibnamefont {Iorsh}}, \bibinfo {author} {\bibfnamefont {A.~V.}\ \bibnamefont {Poshakinskiy}},\ and\ \bibinfo {author} {\bibfnamefont {A.~N.}\ \bibnamefont {Poddubny}},\ }\bibfield  {title} {\bibinfo {title} {Waveguide quantum electrodynamics: Collective radiance and photon-photon correlations},\ }\href {https://doi.org/10.1103/RevModPhys.95.015002} {\bibfield  {journal} {\bibinfo  {journal} {Rev. Mod. Phys.}\ }\textbf {\bibinfo {volume} {95}},\ \bibinfo {pages} {015002} (\bibinfo {year} {2023})}\BibitemShut {NoStop}%
\bibitem [{\citenamefont {Brehm}\ \emph {et~al.}(2021)\citenamefont {Brehm}, \citenamefont {Poddubny}, \citenamefont {Stehli}, \citenamefont {Wolz}, \citenamefont {Rotzinger},\ and\ \citenamefont {Ustinov}}]{Brehm2021NPJQM}%
  \BibitemOpen
  \bibfield  {author} {\bibinfo {author} {\bibfnamefont {J.~D.}\ \bibnamefont {Brehm}}, \bibinfo {author} {\bibfnamefont {A.~N.}\ \bibnamefont {Poddubny}}, \bibinfo {author} {\bibfnamefont {A.}~\bibnamefont {Stehli}}, \bibinfo {author} {\bibfnamefont {T.}~\bibnamefont {Wolz}}, \bibinfo {author} {\bibfnamefont {H.}~\bibnamefont {Rotzinger}},\ and\ \bibinfo {author} {\bibfnamefont {A.~V.}\ \bibnamefont {Ustinov}},\ }\bibfield  {title} {\bibinfo {title} {Waveguide bandgap engineering with an array of superconducting qubits},\ }\href {https://doi.org/10.1038/s41535-021-00310-z} {\bibfield  {journal} {\bibinfo  {journal} {npj Quantum Materials}\ }\textbf {\bibinfo {volume} {6}},\ \bibinfo {pages} {10} (\bibinfo {year} {2021})}\BibitemShut {NoStop}%
\bibitem [{\citenamefont {Liedl}\ \emph {et~al.}(2024)\citenamefont {Liedl}, \citenamefont {Tebbenjohanns}, \citenamefont {Bach}, \citenamefont {Pucher}, \citenamefont {Rauschenbeutel},\ and\ \citenamefont {Schneeweiss}}]{Liedl2024PRX}%
  \BibitemOpen
  \bibfield  {author} {\bibinfo {author} {\bibfnamefont {C.}~\bibnamefont {Liedl}}, \bibinfo {author} {\bibfnamefont {F.}~\bibnamefont {Tebbenjohanns}}, \bibinfo {author} {\bibfnamefont {C.}~\bibnamefont {Bach}}, \bibinfo {author} {\bibfnamefont {S.}~\bibnamefont {Pucher}}, \bibinfo {author} {\bibfnamefont {A.}~\bibnamefont {Rauschenbeutel}},\ and\ \bibinfo {author} {\bibfnamefont {P.}~\bibnamefont {Schneeweiss}},\ }\bibfield  {title} {\bibinfo {title} {Observation of superradiant bursts in a cascaded quantum system},\ }\href {https://doi.org/10.1103/PhysRevX.14.011020} {\bibfield  {journal} {\bibinfo  {journal} {Physical Review X}\ }\textbf {\bibinfo {volume} {14}},\ \bibinfo {pages} {011020} (\bibinfo {year} {2024})}\BibitemShut {NoStop}%
\bibitem [{\citenamefont {Clemens}\ \emph {et~al.}(2003)\citenamefont {Clemens}, \citenamefont {Horvath}, \citenamefont {Sanders},\ and\ \citenamefont {Carmichael}}]{few_atoms_charmichael}%
  \BibitemOpen
  \bibfield  {author} {\bibinfo {author} {\bibfnamefont {J.~P.}\ \bibnamefont {Clemens}}, \bibinfo {author} {\bibfnamefont {L.}~\bibnamefont {Horvath}}, \bibinfo {author} {\bibfnamefont {B.~C.}\ \bibnamefont {Sanders}},\ and\ \bibinfo {author} {\bibfnamefont {H.~J.}\ \bibnamefont {Carmichael}},\ }\bibfield  {title} {\bibinfo {title} {Collective spontaneous emission from a line of atoms},\ }\href {https://doi.org/10.1103/PhysRevA.68.023809} {\bibfield  {journal} {\bibinfo  {journal} {Phys. Rev. A}\ }\textbf {\bibinfo {volume} {68}},\ \bibinfo {pages} {023809} (\bibinfo {year} {2003})}\BibitemShut {NoStop}%
\bibitem [{\citenamefont {Carmichael}\ and\ \citenamefont {Kim}(2000)}]{charmichael_2}%
  \BibitemOpen
  \bibfield  {author} {\bibinfo {author} {\bibfnamefont {H.}~\bibnamefont {Carmichael}}\ and\ \bibinfo {author} {\bibfnamefont {K.}~\bibnamefont {Kim}},\ }\bibfield  {title} {\bibinfo {title} {A quantum trajectory unraveling of the superradiance master equation.},\ }\href {https://doi.org/https://doi.org/10.1016/S0030-4018(99)00694-X} {\bibfield  {journal} {\bibinfo  {journal} {Optics Communications}\ }\textbf {\bibinfo {volume} {179}},\ \bibinfo {pages} {417} (\bibinfo {year} {2000})}\BibitemShut {NoStop}%
\bibitem [{\citenamefont {Gross}\ and\ \citenamefont {Haroche}(1982)}]{gross_haroche}%
  \BibitemOpen
  \bibfield  {author} {\bibinfo {author} {\bibfnamefont {M.}~\bibnamefont {Gross}}\ and\ \bibinfo {author} {\bibfnamefont {S.}~\bibnamefont {Haroche}},\ }\bibfield  {title} {\bibinfo {title} {Superradiance: An essay on the theory of collective spontaneous emission},\ }\href {https://doi.org/https://doi.org/10.1016/0370-1573(82)90102-8} {\bibfield  {journal} {\bibinfo  {journal} {Physics Reports}\ }\textbf {\bibinfo {volume} {93}},\ \bibinfo {pages} {301} (\bibinfo {year} {1982})}\BibitemShut {NoStop}%
\bibitem [{\citenamefont {Carmichael}(2015)}]{Carmichael2015breakdown}%
  \BibitemOpen
  \bibfield  {author} {\bibinfo {author} {\bibfnamefont {H.~J.}\ \bibnamefont {Carmichael}},\ }\bibfield  {title} {\bibinfo {title} {Breakdown of photon blockade: A dissipative quantum phase transition in zero dimensions},\ }\href {https://doi.org/10.1103/PhysRevX.5.031028} {\bibfield  {journal} {\bibinfo  {journal} {Phys. Rev. X}\ }\textbf {\bibinfo {volume} {5}},\ \bibinfo {pages} {031028} (\bibinfo {year} {2015})}\BibitemShut {NoStop}%
\bibitem [{\citenamefont {Beaulieu}\ \emph {et~al.}(2025)\citenamefont {Beaulieu}, \citenamefont {Minganti}, \citenamefont {Frasca}, \citenamefont {Savona}, \citenamefont {Felicetti}, \citenamefont {Candia},\ and\ \citenamefont {Scarlino}}]{Beaulieu2025DPTKerr}%
  \BibitemOpen
  \bibfield  {author} {\bibinfo {author} {\bibfnamefont {G.}~\bibnamefont {Beaulieu}}, \bibinfo {author} {\bibfnamefont {F.}~\bibnamefont {Minganti}}, \bibinfo {author} {\bibfnamefont {S.}~\bibnamefont {Frasca}}, \bibinfo {author} {\bibfnamefont {V.}~\bibnamefont {Savona}}, \bibinfo {author} {\bibfnamefont {S.}~\bibnamefont {Felicetti}}, \bibinfo {author} {\bibfnamefont {R.~D.}\ \bibnamefont {Candia}},\ and\ \bibinfo {author} {\bibfnamefont {P.}~\bibnamefont {Scarlino}},\ }\bibfield  {title} {\bibinfo {title} {Observation of first- and second-order dissipative phase transitions in a two-photon driven kerr resonator},\ }\href {https://doi.org/10.1038/s41467-025-56830-w} {\bibfield  {journal} {\bibinfo  {journal} {Nature Communications}\ }\textbf {\bibinfo {volume} {16}},\ \bibinfo {pages} {1954} (\bibinfo {year} {2025})}\BibitemShut {NoStop}%
\bibitem [{\citenamefont {Janson}(2004)}]{janson2004functional}%
  \BibitemOpen
  \bibfield  {author} {\bibinfo {author} {\bibfnamefont {S.}~\bibnamefont {Janson}},\ }\bibfield  {title} {\bibinfo {title} {Functional limit theorems for multitype branching processes and generalized {P}ólya urns},\ }\href {https://doi.org/https://doi.org/10.1016/j.spa.2003.12.002} {\bibfield  {journal} {\bibinfo  {journal} {Stochastic Processes and their Applications}\ }\textbf {\bibinfo {volume} {110}},\ \bibinfo {pages} {177} (\bibinfo {year} {2004})}\BibitemShut {NoStop}%
\bibitem [{\citenamefont {Fink}\ \emph {et~al.}(2018)\citenamefont {Fink}, \citenamefont {Schade}, \citenamefont {H\"{o}fling}, \citenamefont {Schneider},\ and\ \citenamefont {Imamoglu}}]{Fink2017NatPhys}%
  \BibitemOpen
  \bibfield  {author} {\bibinfo {author} {\bibfnamefont {T.}~\bibnamefont {Fink}}, \bibinfo {author} {\bibfnamefont {A.}~\bibnamefont {Schade}}, \bibinfo {author} {\bibfnamefont {S.}~\bibnamefont {H\"{o}fling}}, \bibinfo {author} {\bibfnamefont {C.}~\bibnamefont {Schneider}},\ and\ \bibinfo {author} {\bibfnamefont {A.}~\bibnamefont {Imamoglu}},\ }\bibfield  {title} {\bibinfo {title} {Signatures of a dissipative phase transition in photon correlation measurements},\ }\href {https://doi.org/10.1038/s41567-017-0020-9} {\bibfield  {journal} {\bibinfo  {journal} {Nature Physics}\ }\textbf {\bibinfo {volume} {14}},\ \bibinfo {pages} {365} (\bibinfo {year} {2018})}\BibitemShut {NoStop}%
\bibitem [{\citenamefont {Minganti}\ \emph {et~al.}(2018)\citenamefont {Minganti}, \citenamefont {Biella}, \citenamefont {Bartolo},\ and\ \citenamefont {Ciuti}}]{minganti2018spectral}%
  \BibitemOpen
  \bibfield  {author} {\bibinfo {author} {\bibfnamefont {F.}~\bibnamefont {Minganti}}, \bibinfo {author} {\bibfnamefont {A.}~\bibnamefont {Biella}}, \bibinfo {author} {\bibfnamefont {N.}~\bibnamefont {Bartolo}},\ and\ \bibinfo {author} {\bibfnamefont {C.}~\bibnamefont {Ciuti}},\ }\bibfield  {title} {\bibinfo {title} {Spectral theory of liouvillians for dissipative phase transitions},\ }\href {https://doi.org/10.1103/PhysRevA.98.042118} {\bibfield  {journal} {\bibinfo  {journal} {Phys. Rev. A}\ }\textbf {\bibinfo {volume} {98}},\ \bibinfo {pages} {042118} (\bibinfo {year} {2018})}\BibitemShut {NoStop}%
\bibitem [{\citenamefont {Sieberer}\ \emph {et~al.}(2016)\citenamefont {Sieberer}, \citenamefont {Buchhold},\ and\ \citenamefont {Diehl}}]{Sieberer2016RPP}%
  \BibitemOpen
  \bibfield  {author} {\bibinfo {author} {\bibfnamefont {L.~M.}\ \bibnamefont {Sieberer}}, \bibinfo {author} {\bibfnamefont {M.}~\bibnamefont {Buchhold}},\ and\ \bibinfo {author} {\bibfnamefont {S.}~\bibnamefont {Diehl}},\ }\bibfield  {title} {\bibinfo {title} {Keldysh field theory for driven open quantum systems: A review},\ }\href {https://doi.org/10.1088/0034-4885/79/9/096001} {\bibfield  {journal} {\bibinfo  {journal} {Rep. Prog. Phys.}\ }\textbf {\bibinfo {volume} {79}},\ \bibinfo {pages} {096001} (\bibinfo {year} {2016})}\BibitemShut {NoStop}%
\bibitem [{\citenamefont {Reilly}\ \emph {et~al.}(2026)\citenamefont {Reilly}, \citenamefont {J{\"a}ger}, \citenamefont {Cooper},\ and\ \citenamefont {Holland}}]{Reilly2026}%
  \BibitemOpen
  \bibfield  {author} {\bibinfo {author} {\bibfnamefont {J.~T.}\ \bibnamefont {Reilly}}, \bibinfo {author} {\bibfnamefont {S.~B.}\ \bibnamefont {J{\"a}ger}}, \bibinfo {author} {\bibfnamefont {J.}~\bibnamefont {Cooper}},\ and\ \bibinfo {author} {\bibfnamefont {M.~J.}\ \bibnamefont {Holland}},\ }\bibfield  {title} {\bibinfo {title} {Fully collective superradiant lasing with vanishing sensitivity to cavity length vibrations},\ }\href {https://doi.org/10.1103/v6jq-m6sk} {\bibfield  {journal} {\bibinfo  {journal} {Physical Review Letters}\ }\textbf {\bibinfo {volume} {136}},\ \bibinfo {pages} {143803} (\bibinfo {year} {2026})}\BibitemShut {NoStop}%
\end{thebibliography}
\end{document}